\documentclass[11pt]{article}

\pdfoutput=1

\usepackage{amsfonts,latexsym,amsthm,amssymb,amsmath,amscd,euscript}
\usepackage{fullpage}
\usepackage[margin=1.0in]{geometry}
\usepackage[dvipsnames]{xcolor}
\usepackage{hyperref}
\hypersetup{
    colorlinks=true,
    linkcolor=OrangeRed,
    citecolor=ForestGreen,
    filecolor=Orchid,
    urlcolor=Cerulean
}
\usepackage{mathtools}
\usepackage{charter}
\usepackage{natbib}

\usepackage{varioref}

\usepackage{cancel}
\usepackage[dvipsnames]{xcolor}
\usepackage[usenames,dvipsnames]{pstricks}

\definecolor{royalpurple}{rgb}{0.47, 0.32, 0.66}

\newcommand{\T}{\mathcal{T}}

\newcommand{\pgc}{P_{\text{GC}}}
\newcommand{\pgd}{P_{\text{GD}}}
\newcommand{\pbc}{P_{\text{BC}}}
\newcommand{\pbd}{P_{\text{BD}}}

\newcommand{\ass}{u_a}

\newcommand{\salary}{s}

\newcommand{\D}{\delta}

\newcommand{\tax}{r}
\newcommand{\Dbribe}{\beta}
\newcommand{\selstr}{\Omega}
\newcommand{\bribesize}{\mathcal{B}}

\title{Institutions of public judgment established \\ by social contract and taxation}
\author{Taylor A. Kessinger and Joshua B. Plotkin}

\begin{document}

\maketitle

\section*{Abstract}

{\small Indirect reciprocity is a plausible mechanism for sustaining cooperation: people cooperate with those who have a good reputation, which can be acquired by helping others.
However, this mechanism requires the population to agree on who has good or bad moral standing.
Consensus can be provided by a central institution that monitors and broadcasts reputations.
But how might such an institution be maintained, and how can a population ensure that it is effective and incorruptible?
Here we explore a simple mechanism to sustain an institution of reputational judgment: a compulsory contribution from each member of the population, i.e., a tax.
We analyze the maximum possible tax rate that individuals will rationally pay to sustain an institution of judgment, which provides a public good in the form of information, and we derive necessary conditions for individuals to resist the temptation to evade their tax payment.
We also consider the possibility that institution members may be corrupt and subject to bribery, and we analyze how often an institution must be audited to prevent bribery.
Our analysis has implications for the establishment of robust public institutions that provide social information to support cooperation in large populations---and the potential negative consequences associated with wealth or income inequality.}

\setcounter{section}{1}
\section*{Introduction}

Humans are among the most social organisms on the planet.
Our societies depend on high levels of cooperation, whether through direct acts of altruism, contributions to the public good, or forgoing opportunities to harm others.
But prosocial behavior frequently incurs a cost to the individual, so there is a strong incentive for self-interested individuals \emph{not} to cooperate.
To make matters worse, healthy societies often demand cooperation between large numbers of genetically unrelated individuals, who may have no history of past, or prospect for future, interactions -- so that mechanisms such as kin selection and direct reciprocity (tit-for-tat) cannot explain widespread cooperative behavior in large human societies.

In order for cooperation to proliferate, its benefit must disproportionately accrue to those who cooperate. Reputations and moral standing can facilitate this: individuals who behave prosocially may be rewarded with a good reputation and, thus, may be targeted for cooperative acts  in the future.
Via this mechanism, known as ``indirect reciprocity,'' a prosocial act is rewarded not by the recipient, but by a third party who observes the interaction or learns of it.
When players condition their cooperation on the moral standing of their co-player, a high rate of overall cooperation can emerge, as those of good standing are likely to reward each other and to punish those of bad standing.

Indirect reciprocity is, \emph{prima facie}, a plausible and realistic mechanism for sustaining cooperation in large populations.
Humans are known to care about their own reputations, as well as the reputations of those they interact with.
There is abundant empirical evidence that humans consider reputations and moral standing when deciding how to treat others \citep{alexander_biology_1987,elster_social_1989,cialdini_focus_1991,kandori_social_1992,fehr_social_2004,bicchieri_grammar_2005, tomasello_origins_2013, fehr_normative_2018, hechter_norms_2018, curry_good_2019}.
Even toddlers selectively cooperate with those they have observed behaving prosocially in the past \citep{nava2019}.

In standard models of indirect reciprocity, players rely on a rule known as a ``social norm'' for assigning reputations.
Foundational work on indirect reciprocity \citep{ohtsuki_goodness_2004, ohtsuki_leading_2006} has identified a ``leading eight'' set of social norms that can sustain a high rate of cooperation.
However, most of these norms suffice only when reputations are matters of societal consensus, so that everyone agrees on who is ``good'' or ``bad'' at any point in time.
In the absence of such consensus, prosocial behavior by one player (for example, cooperation directed toward someone that player views as good) may be interpreted as antisocial behavior by an observer (because the observer views the recipient as bad), which can lead to a collapse of reputations and eventual loss of cooperation \citep{hilbe_indirect_2018}.

Most models of indirect reciprocity do not trouble themselves with the details of how agreement about reputations arises; rather, they take it for granted that consensus reputations exist in the population.
Exactly \emph{how} populations come to agree about reputations remains an under-studied topic.
Several plausible mechanisms can yield such consensus.
For example, rapid gossip about reputations can ensuring sufficient consensus to sustain cooperation \citep{sommerfeld_gossip_2007,kawakatsu_mechanistic_2024}.
Alternately, empathetic perspective-taking can also effectively ensure agreement \citep{radzvilavicius_empathy_2019}.

Here we consider an alternative mechanism that is robustly and widely employed by human societies: an institution of judgment that assigns and broadcasts reputational judgments \citep{radzvilavicius_adherence_2021}.
By furnishing consensus about reputations, an institution can dramatically increase the level of cooperation in a population.
In doing so, it can resolve conflict between individuals or even groups that arise as a result of conflicting needs or incompatible reputational views---aiding the transition from privatized, retributive modes of justice to formalized, bureaucratic ones \citep{blackmichaud_feuding_1975,fukuyama_origins_2011,posner_economics_1981,boehm_hierarchy_1999,gluckman_politics_1965}.
For example, under the social norm ``Stern Judging'', punishing bad individuals is mandatory.
When reputations are privately held, disagreement about who is ``good'' and ``bad'' means it is impossible for anyone to maintain a good reputation in the eyes of the majority of the population \citep{uchida_effect_2013, oishi_group_2013,traag_dynamical_2013}.
Accordingly, the optimal strategy is unconditional defection, and the expected rate of cooperation is $0\%$.
If an institution relieves disagreement by broadcasting consensus reputational judgments, the rate of cooperation can be very high---close to $100\%$, if mistakes are rare \citep{pacheco_stern-judging_2006}.

Indirect reciprocity can solve the second-order free-rider problem, because the task of punishing ``bad'' individuals is shared by the population at large and may itself be rewarded with a good reputation \citep{panchanathan_indirect_2004, sasaki_indirect_2016}.
But mechanisms to ensure consensus about reputations may in turn carry their own costs; thus, a different sort of free-rider problem persists.
Gossip may be time-consuming, for example, and empathetic perspective-taking can be cognitively demanding.
In the case of an institution, performing and broadcasting judgments requires effort---a cost paid by institution members.
At the same time, an institution provides public information that raises the rate of cooperation and increases the overall welfare of society.
And so institution members may demand, and members of a society may be willing to furnish, a common pool resource for maintaining the institution and supporting the public (informational) benefit it provides.
We refer to such a common pool resource as a ``tax'' (though we note that the ``tax'' we discuss here is not necessarily monetary in nature; any kind of mandatory contribution levied upon a population for the purpose of sustaining an institution would qualify).

In this paper, we use classical models from evolutionary game theory and adaptive dynamics to study how taxation can support an institution of moral judgment.
We compute the maximum tax rate that rational individuals will willingly pay to support such an institution.
Because the tax represents an individual cost, individuals may be tempted not to pay it, which would starve the institution of required funding.
Hence, we analyze a plausible way to deter individuals from evading their tax obligations: punishing them with a bad reputation.
We show that this is a simple and effective way to safeguard against tax evasion---and one that simultaneously solves the second-order free rider problem.
We determine exactly how often members of the population need to be audited to ensure that tax dodgers do not proliferate, in terms of the costs and benefits of cooperation, the tax rate, the institution size and strictness, and error rates.

Finally, institution members may themselves be tempted to augment their income by soliciting a bribe---i.e. a side-payment by a member of the population, in return for being assigned a good reputation regardless of their behavior.
We show that periodic auditing of institution members can stabilize cooperation against invasion by individuals who bribe to receive good reputations.
We determine exactly how frequently such audits must take place in order to prevent invasion by bribers and subsequent loss of cooperation.

Our model differs markedly from previous studies of the effect of institutions on cooperation, such as \citet{chiba-okabe_2024} and \citet{lie-panis_social_2023}.
In both of those models, the institution plays an active role in redistributing wealth from cooperators to defectors; in the latter, institutions also increase the likelihood that a player's reputation is known (this would be equivalent, in our model, to reducing the rate of assessment error). The institution we study here is  simpler: institutions cannot directly reward cooperation, punish defection, or reduce the uncertainty associated with someone's reputation.
Their \emph{sole} function is to facilitate consensus by providing a reputation broadcast.
Despite this very limited role for the institution in our model, it can still dramatically alter the cooperative landscape. And the institution we study can be seen as a minimal model for consensus-building processes––and how much players will willingly pay to maintain this consensus––in indirect reciprocity.

\section*{Model}

We consider a large population of size $N$; each round, every individual plays a donation game with everyone else, acting once as a donor and once as a recipient.
Donors who cooperate pay a cost $c>0$ to convey a benefit $b > c$ to their interaction partner; donors who defect pay no cost and convey no benefit.
Each donor acts according to their behavioral strategy.
We consider two major classes of behavioral strategy here: \emph{discriminators} (abbreviated DISC) who cooperate with individuals they see as ``good'' and defect with those they see as ``bad'', and \emph{defectors} (abbreviated ALLD) who defect with everyone.
As game play occurs, each individual $i$ accrues a strategy-dependent payoff $\Pi_i$ from their social interactions. We denote the average payoff to a player using strategy $s \in \{ \text{ALLD},\text{DISC} \}$ as $\Pi_s$.
Strategies then evolve via social learning: individuals preferentially copy the strategies of those with higher payoffs (see Methods).

The labels ``good'' and ``bad'' (which we refer to as moral standing or reputation) are assigned by an institution \citep{radzvilavicius_adherence_2021}.
An institution consists of $Q$ members and is characterized by a strictness threshold $q$.
To assess an individual's reputation, an institution member assesses a random interaction in the previous round in which that individual acted as a donor. They consider the donor's action, accounting for the recipient's reputation, using a rule called a \emph{social norm}.
With probability $0 < \ass < 1/2$, an institution member may err in their assessment (i.e., assess a bad individual as good or vice versa).
This is comparable to the error model outlined in \citet{nowak_evolution_1998}.
Finally, if at least $q$ institution members assess an individual as good, their reputation is broadcast to the entire population as good; otherwise, it is broadcast as bad \citep{radzvilavicius_adherence_2021}.

We focus here on the social norm Stern Judging, which assesses a donor as good if they cooperate with a good recipient or defect with a bad recipient, and assesses them as bad otherwise.
Stern Judging is one of the ``leading eight'' social norms \citep{ohtsuki_goodness_2004,ohtsuki_leading_2006}, which sustains full and stable cooperation in a population provided reputations are publicly shared; however, it cannot sustain cooperation when individuals rely exclusively on private judgments \citep{uchida_effect_2013,oishi_group_2013,traag_dynamical_2013,hilbe_indirect_2018,okada_solution_2018}.

Our primary focus is on the rate and stability of cooperation once an institution of moral judgment is established.
The \emph{de novo} establishment of such an institution is a problem that rightly belongs to political science and thus is outside the scope of our analysis.
We say that cooperation is stable provided a population of discriminators can resist invasion by a rare defector mutant, i.e., the defector strategy cannot spread via social contagion.
This is tantamount to requiring that, in a population of discriminators, the payoff to a discriminator exceeds that of a very rare defector: i.e., $\Pi_{\text{DISC}} > \Pi_{\text{ALLD}}$.

\section*{Results}

We first review the evolutionary game-theoretic conditions under which indirect reciprocity can sustain cooperation, when reputations are public knowledge.
Then, we outline the challenge of maintaining cooperation when players assess reputations based on their private views of each other, a situation we refer to as ``private assessment''.
This comparison quantifies the advantage that an institution of public assessment provides to a population, in terms of the average rate of cooperation and mean payoff in the population.
Because of the payoff advantage an institution provides, rational individuals will be willing to pay some amount to support such an institution each round, which we refer to as a tax.
We derive the maximum possible tax rate a population of rational individuals will willingly pay to support an institution that broadcasts reputations.
We then analyze a method to suppress tax dodgers by auditing them and assigning them bad reputations. 
Finally, we round out our results by establishing conditions under which the strategy of bribing institution members in return for a good reputation will fail to spread by social contagion.

When all individuals in a population are discriminators, they cooperate at rate $G_{\text{DISC}}$, which is precisely the fraction of individuals an arbitrary observer deems to have a good reputation.
Their total fitness is thus given by $\Pi_{\text{DISC}} = (b - c) G_{\text{DISC}}$.
Defectors, on the other hand, do not pay the cost of cooperation, and they accrue a payoff from their interactions with discriminators who view them as good; thus, their fitness is given by $\Pi_{\text{ALLD}} = b G_{\text{ALLD}}$.
Setting $f_{\text{DISC}} = 1$ and solving for when $\Pi_{\text{DISC}} > \Pi_{\text{ALLD}}$ provides the following result: a population of discriminators is stable against invasion by rare defector mutants provided the benefit-to-cost ratio $b/c$ exceeds a critical value $(b/c)^*$, given by
\begin{equation}
    \left( \frac{b}{c} \right)^* =  \begin{cases} 
        1 + \dfrac{G_{\text{ALLD}}}{G_{\text{DISC}} - G_{\text{ALLD}}}\Big|_{f_{\text{DISC}} = 1} & \text{ when } G_{\text{DISC}}|_{f_{\text{DISC}} = 1} > G_{\text{ALLD}}|_{f_{\text{DISC}} = 1}\\
        \infty & \text{ when } G_{\text{DISC}}|_{f_{\text{DISC}} = 1} \leq G_{\text{ALLD}}|_{f_{\text{DISC}} = 1}.
    \end{cases}
    \label{eq:bccond_general}
\end{equation}
The stability of cooperation under indirect reciprocity thus depends on discriminators cooperating with each other more than they do with rare defectors, so that $G_{\text{DISC}}|_{f_{\text{DISC}} = 1} > G_{\text{ALLD}}|_{f_{\text{DISC}} = 1}$.
Provided this condition is satisfied, there is some finite benefit-to-cost ratio $b/c$ for which cooperation is stable.

Under private assessment, individuals rely not on commonly shared consensus reputations but rather on their own private judgments.
Private judgments of a focal individual's standing may not agree, due to independently sampled observations of said individual as well as independent errors in reputation assignment.
The full form of $G_{\text{DISC}}$ under private assessment is given in Equation \ref{eq:reps_private} (see Methods); we briefly outline why it is difficult to maintain a good reputation under private assessment.
A discriminator may cooperate with an individual whom they see as good but whom an observer sees as bad.
Likewise, they may defect with an individual they see as bad but an observer sees as good.
Under the social norms \emph{Shunning} and \emph{Stern Judging}, either of these actions will lead to the focal individual being assigned a bad reputation.
In particular, under \emph{Stern Judging},
an individual's average reputation (averaged over all observers) is $1/2$ no matter what their behavioral strategy is \citep{uchida_effect_2013}.
That is, at any given time, half the population will view an arbitrary individual as good, and half will view them as bad, irrespective of their behavioral strategy.
As a result, the denominator in Equation \ref{eq:bccond_general} goes to zero;
defectors (who never pay the cost of cooperation) are always fitter than discriminators (who pay the cost of cooperation half the time), so the all-discriminator equilibrium is unstable.
Rare defector mutants will drag the population toward the sole stable equilibrium, in which everyone is a defector.
We thus have
\begin{equation}
    \Pi_\text{private} = 0,
\end{equation}
i.e., when the population relies on private assessment, the average payoff is zero.

The inevitability of defectors invading the population is a direct byproduct of the fact that, under private assessment, players' opinions of each other are essentially independent, meaning there is a high risk of disagreement about reputations.
Public information solves this problem by allowing the entire population to agree about each individual's reputation.
Thus, discriminators who cooperate with someone they see as good (or, under Stern Judging, defect with someone they see as bad) can be assured the rest of the population shares their view and thus assigns them a good reputation.
Cooperation mediated by reputation is therefore more much stable when reputations are public information--e.g., monitored and broadcast by some form of public institution.
As a result, it is possible to satisfy Equation \ref{eq:bccond_general}, meaning there is a stable equilibrium where everyone is a discriminator.
At this equilibrium, individuals cooperate with each other at rate $G_{\text{DISC}}$ and thus gather an average payoff
\begin{equation}
    \Pi_\text{institution} = (b - c) G_{\text{DISC}}|_{f_{\text{DISC}} = 1}.
\end{equation}
For example, in the simple case of \emph{Stern Judging} and $Q = 1$ (i.e., the institution has a single member), we have $G_{\text{DISC}}|_{f_{\text{DISC}} = 1} = 1 - \ass$ and $G_{\text{ALLD}}|_{f_{\text{DISC}} = 1} = 2\ass(1 - \ass)$; discriminators cooperate with each other a fraction $1 - \ass$ of the time, and defectors can only be assigned good reputations either by accident or by defecting against a discriminator who was themselves erroneously assigned a bad reputation.
Accordingly, the critical ratio $(b/c)^*$ is
\begin{equation}
    \left( \frac{b}{c} \right)^* = 1 + \frac{2 \ass}{1 - 2\ass},
\end{equation}
which, for small $\ass$, is barely greater than $1$. That is, the condition to stabilize cooperation in this case is only slightly more stringent than the condition needed for the donation game to be a prisoner's dilemma in the first place.

\subsection*{Maximum rational tax}
Under the social norms we consider here, a stable equilibrium ($f_{\text{DISC}} = 1$) with a high rate of cooperation arises only in the presence of an institution that provides consensus information about reputations.
In other words, the difference between $\Pi_\text{institution}$ and $\Pi_\text{private}$ is the value created by the existence of an institution.
As a result, members of the institution may demand, and individuals in the population will willingly pay, a population-wide mandatory fee---i.e., a tax---to sustain the operation of the institution.
The maximum possible tax rate is
\begin{equation}
    \mathcal{T} = \Pi_\text{institution} - \Pi_\text{private} = (b - c) G_{\text{DISC}}|_{f_{\text{DISC}} = 1}.
\end{equation}
As an example, when the institution consists of only a single member ($Q = 1$) and the norm is Stern Judging, we have
\begin{equation}
    \mathcal{T} = \Pi_\text{institution} - \Pi_\text{private} = (b - c) (1 - \ass).
\end{equation}
Individuals are incentivized to pay any tax up to this amount and are indifferent to paying this full amount; the latter yields the same fitness (zero) as a population of defectors, which would occur in the absence of an institution.
If individuals pay a fraction $\tax$ of the theoretical maximum tax, the fitness of discriminators becomes
\begin{equation}
        \Pi_{\text{DISC}}^\prime = (b - c) G_{\text{DISC}} - \tax \mathcal{T} = (1 - \tax) (b - c) G_{\text{DISC}}.
\end{equation}
Summing this amount over the $N$ individuals in a population yields a total revenue $N \tax \mathcal{T}$ per round; dividing this evenly among the $Q$ members of the institution yields a salary
\begin{equation}
    \salary = N \tax \mathcal{T} / Q = N \tax (b - c) G_{\text{DISC}} / Q.
    \label{eq:salary}
\end{equation}
per round for each member of the institution.
For example, under Stern Judging and with $Q = 1$, this salary is
\begin{equation}
    \salary = N \tax (b - c) (1 - \ass).
    \label{eq:SJ_salary}
\end{equation}
Institution members' salaries are thus linked to the societal welfare they provide; their salary increases with the net payoff to discriminators (the cooperation rate times the fitness each individual accrues as a result of their interactions), whereas the tax rate $\tax$ \emph{increases} the salary of institution members but \emph{decreases} the fitness of discriminators.
As we will shortly see, this can have complicated effects on the survival of institutions.

\subsection*{Protection against tax evaders}

Taxation imposes a cost on individuals and thus reduces their fitness (but not to zero).
As a result, individuals could increase their fitness by shirking their tax obligations.
We consider an individual's tax status---that is, their decision whether or not to pay the tax---as an additional aspect of their behavioral strategy.
If tax evasion increases a player's fitness, it can spread via social contagion.
One way to deter tax evasion is for the institution to periodically audit individuals (as, indeed, governments regularly do) to determine whether they have paid the expected tax.
We assume the institution performs an audit each round with probability $\D$---meaning that an individual who evades the tax is discovered with this probability.
Once a tax evader is discovered, the institution will ``punish'' them by assigning them a bad public reputation; the population of discriminators will then ``carry out" the punishment by defecting against that individual.
(The punishment is therefore cost-free to the population, which avoids the second-order free-rider problem.)

By starving an institution of funding, tax cheats can potentially destabilize cooperation; if they become too numerous, the institution cannot perform its function properly.
Institutions have an incentive to prevent this (lest they lose their funding), as do societies at large---because, in the absence of an institution, mutual defection is assured.
We have proven that tax-evading \emph{discriminators} cannot invade provided $\D > \tax$, a result that is consistent with
\citet{sasaki_indirect_2016}.
This result is intuitive: the steeper the tax rate $\tax$, the more an individual benefits by not paying it.
To compensate for this, potential tax cheats must be subjected to a greater risk of discovery, $\D$, to suppress their spread.

For the remainder of our analysis, we consider tax-evading \emph{defectors}, for whom reputations and fitnesses are given by
\begin{equation}
    \begin{split}
        G_{\text{ALLD}}^\prime & = (1 - \D) G_{\text{ALLD}},
        \\
        \Pi_{\text{ALLD}}^\prime & = b G_{\text{ALLD}}^\prime.
    \end{split}
\end{equation}
Our first main result is that tax-evading defectors will fail to invade the population provided the benefit-to-cost ratio exceeds the following critical value:
\begin{equation}
    \left( \dfrac{b}{c} \right)^* =   
    \begin{cases}
        1 + \dfrac{(1 - \D)G_{\text{ALLD}}}{(1 - \tax) G_{\text{DISC}} - (1 - \D) G_{\text{ALLD}}} & \text{~for~} (1 - \tax) G_{\text{DISC}} > (1 - \D) G_{\text{ALLD}},
        \\
        \infty & \text{~for~} (1 - \tax) G_{\text{DISC}} < (1 - \D) G_{\text{ALLD}}
    \end{cases}
    \label{eq:naive_resistance_cond}
\end{equation}

The cost-benefit ratio required for discriminators to resist invasion by a defector who shirks their tax obligation (Eq.~\ref{eq:naive_resistance_cond}) can be compared to the corresponding classical condition for stability against a defector without tax collection (Eq.~\ref{eq:bccond_general}).
The requirement on $b/c$ to suppress tax evading defectors is generally more stringent than for an institution that is assumed (unrealistically) to operate without any tax revenue at all.
Moreover, Eq.~\ref{eq:naive_resistance_cond} and Eq.~\ref{eq:bccond_general} have almost identical forms except for the prefactors that modify $G_{\text{DISC}}$ and $G_{\text{ALLD}}$; these prefactors reflect that, in the presence of a tax-collecting institution, a discriminator's fitness is reduced by a factor $\tax$ (due to paying the tax), whereas an evader-defector's fitness---which arises from donations by discriminators who see them as good---is reduced by a factor $\D$ (the probability that they are detected shirking the tax). 

The comparison between Eq.~\ref{eq:naive_resistance_cond} (in the context of tax-evading discriminators) and Eq.~\ref{eq:bccond_general} (in the context of no taxation) reveals another important point: in some conditions a tax-evading defector will always be able to invade, regardless of the ratio $b/c$.
This situation cannot occur if an institution operates without taxation, but it can occur in the presence of taxation, namely whenever $(1 - \tax) G_{\text{DISC}} < (1 - \D) G_{\text{ALLD}}$.
In other words, when either the tax rate $\tax$ is too large \emph{or} the chance of detecting a tax evader $\D$ is too small, an evader-defector can invade, regardless of the benefit of cooperation. Eq.~\ref{eq:naive_resistance_cond} therefore provides quantitative guidance on what tax rates and auditing rates are required to permit stable cooperation in the face of defecting tax-dodgers.

Another simple way to understand the relationship between $(b/c)^*$, the tax rate $\tax$, and the detection probability $\D$ is to evaluate the partial derivatives of the payoff difference $\Pi_{\text{DISC}} - \Pi_{\text{ALLD}}$:
\begin{equation}
    \begin{split}
        \partial_{\tax} (\Pi^\prime_{\text{DISC}} - \Pi^\prime_{\text{ALLD}}) & = - (b - c) G_{\text{DISC}}<0,
        \\
        \partial_{\D} (\Pi^\prime_{\text{DISC}} - \Pi^\prime_{\text{ALLD}}) & = b G_{\text{ALLD}}>0.
    \end{split}
\end{equation}
These calculations support a natural intuition: cooperation can be stabilized more easily against tax-evading defectors either by reducing the tax rate or by increasing the chance of detecting tax evaders.

\begin{figure}
    \centering
    \includegraphics[width=0.49\textwidth]{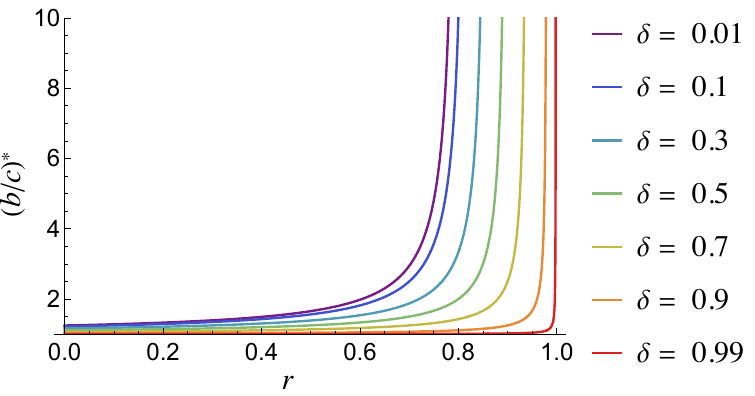}
    \includegraphics[width=0.49\textwidth]{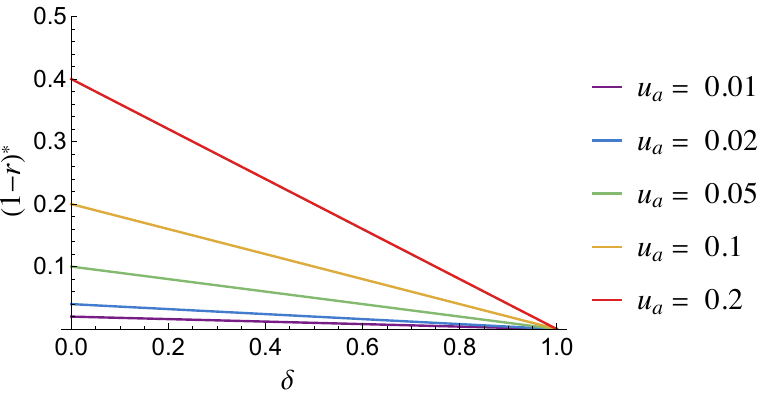}
    \caption{\small{Invasibility of tax-paying discriminators by tax-evading defectors.
    Left:
    the critical cost-benefit ratio $(b/c)^*$ required to resist invasion, as a function of the tax rate $\tax$, for several different values of the chance $\D$ that tax evaders are successfully detected.
    When the detection probability is high, the critical ratio barely exceeds $1$, even for a very large tax rate.
    When detection is rare, however, tax evading defectors can more easily destabilize cooperation (larger value of $(b/c)^*$).
    In fact, when detection is too rare, there is no finite benefit of cooperation whatsoever that resists invasion by tax-evading defectors, i.e., $(b/c)^*=\infty$.
    Right: for each detection rate $\D$, there is a value of $(1 - r) \in [0,1]$ above which it is possible to sustain cooperation, i.e., for which there is a finite, positive critical ratio $(b/c)^*$.
    Below this line, no such value of $(b/c)^*$ exists.
    Similar results hold for larger institutions under \emph{Stern Judging} and for sufficiently tolerant \emph{Shunning} institutions.
    }}
    \label{fig:bc_tax_rate_nobribe}
\end{figure}

As an example of the result above, under Stern Judging and with an institution of size $Q = 1$, the condition for discriminators to resist invasion by defector-evaders (Eq.~\ref{eq:naive_resistance_cond}) simplifies to 
\begin{equation}
    \left( \frac{b}{c} \right)^* =
    \begin{cases}
        1 + \dfrac{2\ass(1 - \D)}{1 - \tax - 2 \ass (1 - \D)} & \text{~when~} (1 - \tax) > 2 \ass (1 - \D) ,
        \\
        \infty & \text{~when~} (1 - \tax) < 2 \ass (1 - \D).
    \end{cases}
\end{equation}
This result is depicted in Figure \ref{fig:bc_tax_rate_nobribe}, for a range of tax rates, evader detection rates, and assessment error rates.

\subsection*{Protection against institutional corruption}

Aside from members of the population at large, who may be motivated to evade paying the tax, institution members themselves (insofar as they are human) may be motivated to increase their payoff beyond their salary.
Likewise, those who stand to be tarnished with a bad reputation, such as tax-evading defectors, may consider an alternative method to ensure a good reputation---namely, to bribe institution members in exchange for a good reputation, in lieu of making a tax payment.
This can result in individuals being assigned a good reputation despite their defecting behavior or status as a tax cheat.
We consider two types of bribing strategies.
First, individuals may simply pay off the institution every round in exchange for a good reputation; we refer to this as \emph{unconditional} bribing.
Second, individuals may pay off the institution \emph{only} if they are first discovered as a tax cheat; we refer to this as \emph{conditional} bribing.

To safeguard against corruption, societies may provide a disincentive for institution members to accept bribes.
One solution is for institution members to undergo periodic auditing to ensure they have not taken bribes.
In our model, institution members who accept a bribe are discovered with probability $\Dbribe$; discovery means they are removed from their position and replaced with someone else, thus forgoing the entirety of their salary. 
(Results are similar if an institution member who is discovered also has their bribe confiscated.\footnote{If bribes are confiscated, this increases the required bribe by a factor $1/(1-\Dbribe)$, but $\Dbribe$ is typically close to zero whereas $N$ is large, so this factor has a small effect.})
Under these conditions, an institution member will accept a bribe only if doing so increases their expected income.
If they do not accept a bribe, their expected income is $\salary$.
If they do, then with probability $\Dbribe$ they are caught accepting the bribe, and their income for the round is $\bribesize$; with probability $1 - \Dbribe$, they are not caught, and their income is $\bribesize + \salary$.
Thus, the minimum bribe they will be willing to accept satisfies
\begin{equation}
    (1 - \Dbribe)(\bribesize + \salary) + \Dbribe \bribesize > \salary,
    \label{eq:bribe_derivation}
\end{equation}
which implies that
\begin{equation}
    \bribesize = \Dbribe \salary = \Dbribe N \tax \mathcal{T}/Q.
\end{equation}
(We assume institution members are concerned only with their payoff in the current round, but relaxing this assumption does not affect the argument; if the institution member intends to continue taking bribes, incorporating time discounting will equally affect both sides of Eq.~ \ref{eq:bribe_derivation}.)

Like the salary itself (Equation \ref{eq:salary}), the bribe size $\bribesize$ required to tempt an institution member is related to the population's total welfare via the average discriminator payoff $(b-c) G_{\text{DISC}}$, and it scales with the tax rate $\tax$.
The tax rate therefore introduces an additional complication: it lowers the fitness of each taxpaying discriminator and thus makes alternative strategies, such as tax-evading defection, more appealing, but it also ensures that bribing defectors must pay a larger bribe to compensate for the risk undertaken by institution members in accepting bribes.
In this way, it can \emph{lower} the fitness of defectors and thus make cooperation more robust.
The details depend on how, and when, a player bribes.
We now consider bribery as an additional element of a player's behavioral strategy, and we analyze when a population is stable against invasion of different types of bribing tax evaders.

\subsubsection*{Importance of institutional auditing}

As we will shortly see, the success of a bribing strategy, and in particular how that success depends on the tax rate $\tax$, is a function of the compound parameter $N \Dbribe$.
Because a single individual is chosen to consider updating their strategy each round of game play, an arbitrary individual is selected for potential updating in a given round with probability $1/N$.
Likewise, an institution member, if corrupt, is identified with probability $\Dbribe$.
The compound term $N \Dbribe$ therefore characterizes the difference in time scales between individual strategy updating (which occurs each round at rate $1/N$) and anti-corruption auditing of institution members (which occurs each round at rate $\Dbribe$).
The stability of cooperation thus turns sensitively on which of these two processes occurs more quickly.
When $N \Dbribe > 1$, institutional auditing is fast relative to strategic updating; institution members who accept bribes are likely to be discovered and replaced more quickly than bribing strategies can spread through the population via social contagion.
By contrast, when $N \Dbribe < 1$, behavioral strategies spread more quickly than auditing occurs, so there is a chance for bribery to destabilize cooperation before corrupt institution members can be detected and removed.

\subsubsection*{Protection against unconditional bribery}

We first consider an ``unconditional'' briber, who always pays the full bribe amount to institution members; this amount is precisely sufficient to incentivize them despite the risk $\Dbribe$ of being caught during an audit.
Such a briber will always be assigned a good reputation, but because they bribe the institution every round, their fitness is reduced by $Q \bribesize = Q \Dbribe \salary$.
Their reputation and fitness are therefore
\begin{equation}
    \begin{split}
        G^\prime_{\text{ALLD}} & = 1
        \\
        \Pi^\prime_{\text{ALLD}} & = b G^\prime_{\text{ALLD}} - Q \Dbribe \salary = b - N \Dbribe \tax ( b - c ) G_{\text{DISC}}.
    \end{split}
\end{equation}
A population of taxpaying discriminators can resist invasion provided $\Pi^\prime_{\text{DISC}} > \Pi^\prime_{\text{ALLD}}$.
We solve this inequality to obtain the minimum benefit-cost ratio required to stabilize cooperation against invasion by defecting tax evaders who bribe unconditionally:
\begin{equation}
    \left( \frac{b}{c} \right)^* =
    \begin{cases}
        1 + \dfrac{1}{G_{\text{DISC}} [1 + (N \Dbribe - 1) \tax] - 1} & \text{ when } N \Dbribe > 1 + \dfrac{1}{\tax} \left( \dfrac{1}{G_{\text{DISC}}} - 1\right)
        \\
        \infty & \text{ when } N \Dbribe \leq 1 + \dfrac{1}{\tax} \left( \dfrac{1}{G_{\text{DISC}}} - 1\right).
    \end{cases}
    \label{eq:unconditional_resistance_cond}
\end{equation}
Note that the chance of detection, $\D$, does not appear in the condition for $(b/c)^*$, because an unconditional briber pays the full bribe amount regardless of whether they are detected.

We have previously argued that the relationship between $\tax$ and $(b/c)^*$ depends on the quantity $N \Dbribe$---which compares the timescale of strategic updates to the timescale of institutional audits.
The result above shows that, for a sufficiently small value of $N \Dbribe$ (i.e., rare auditing), it is impossible to sustain cooperation at all, regardless of the benefit ratio $b/c$; the bribe amount is simply too small, and bribery too tempting, to resist invasion by defector-evader-bribers.
``Sufficiently small'' is determined by the threshold $1 + (1/G_{\text{DISC}} - 1)/\tax$.
A low tax rate (small $\tax$) raises this threshold.
Put another way, when institutions are starved of funding, it is too easy to bribe them, and thus it is impossible to sustain cooperation.
This is especially true when $G_{\text{DISC}} \ll 1$;
the benefit of behaving in accordance with the institution is too weak compared to simply bribing one's way into having a good reputation.
Thus, for stable cooperation to be possible at all, the compound parameter $N \Dbribe$ must exceed $1 + (1/G_{\text{DISC}} - 1)/\tax$.
As a result, a higher tax rate \emph{decreases} the threshold value of $b/c$ required to sustain cooperation---by effectively increasing the amount that tax-evading defectors must pay to be assured a good reputation.

The relationship between the tax rate $\tax$, the compound parameter $N \Dbribe$, the tax-evasion detection probability $\D$, and the stability of cooperation can also be understood by inspecting the partial derivatives 
\begin{equation}
    \begin{split}
        \partial_{\tax} (\Pi^\prime_{\text{DISC}} - \Pi^\prime_{\text{ALLD}}) & = (b - c) G_{\text{DISC}} (N \Dbribe - 1),
        \\
        \partial_{\D} (\Pi^\prime_{\text{DISC}} - \Pi^\prime_{\text{ALLD}}) & = 0.
    \end{split}
\end{equation}
This means that the detection probability $\D$ has no influence on the stability of cooperation, whereas a larger tax rate $\tax$  widens the fitness gap between tax-paying discriminators and tax-evading bribers, making cooperation easier to sustain.
Recall that cooperation can be sustained (i.e. the threshold $(b/c)^*$ is finite) only when $N \Dbribe - 1 > (1/G_{\text{DISC}} - 1) / \tax$; the partial derivative with respect to $\tax$ is positive in this regime.

\begin{figure}
    \centering
    \includegraphics[width=0.49\textwidth]{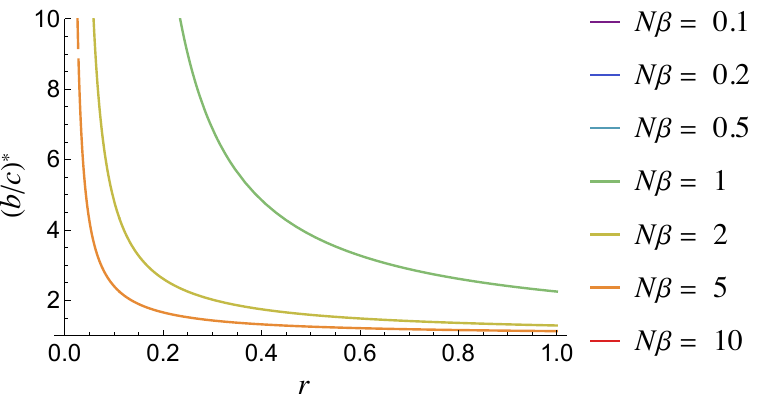}
    \includegraphics[width=0.49\textwidth]{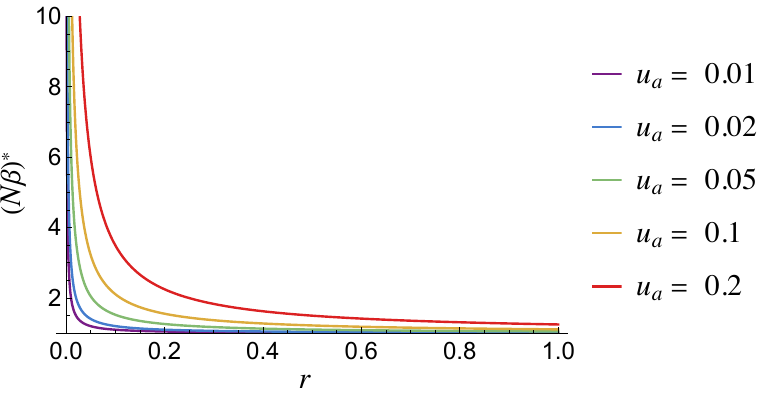}
    \caption{
    \small{
    Invasibility of taxpaying discriminators by  defector-evaders who bribe unconditionally.
    Left: the critical ratio $(b/c)^*$ to ensure DISC is stable against invasion by defecting tax evaders who always bribe institution members, for an institution of size $Q = 1$ that adheres to \emph{Stern Judging}.
    The figure illustrates how this critical ratio depends on the tax rate $\tax$, for several different values of the compound parameter $N \Dbribe$--which determines whether strategic updating or institutional auditing happen more quickly.
    When the tax rate is smaller, it is harder to stabilize DISC, because the salary of institution members is smaller, and therefore it is easier for unconditionally bribing defector-evaders to survive.
    Curves not shown correspond to values of $N \Dbribe$ for which $(b/c)^*=\infty$.
    Right: the critical value of $N \Dbribe$, above which it is possible for taxpaying discriminators to stabilize cooperation for some finite value of $b/c$.
    For $N \Dbribe < 1 + \ass/[\tax(1 - \ass)]$, there is no finite value of $b/c$ for which DISC is stable against such defector-evaders, because auditing is too rare, or institutions are too starved of funding, to stop corruption.
    Similar results hold for larger institutions under \emph{Stern Judging}.
    }
    }
    \label{fig:bc_tax_rate_unconditional}
\end{figure}

As a concrete example, we again consider an institution of size $Q = 1$ under \emph{Stern Judging}; results on invasibility by defecting tax evaders who bribe are presented in figure \ref{fig:bc_tax_rate_unconditional}.
The condition for stable cooperation becomes
\begin{equation}
    \left( \frac{b}{c} \right)^* = 
    \begin{cases}
        1 + \dfrac{1}{(1 - \ass) [1 + (N \Dbribe - 1) \tax] - 1} & \text{ when } N \Dbribe  > 1 + \dfrac{\ass}{\tax (1 - \ass)} \\
         \infty & \text{ when }  N \Dbribe \leq 1 + \dfrac{\ass}{\tax (1 - \ass)}
    \end{cases}
\end{equation}
This exemplifies the fact that a higher tax rate always makes cooperation easier to sustain against such invaders.
This is because the factor $(N \Dbribe - 1)$ that modifies $\tax$ in the denominator is guaranteed to be positive; if it is ever negative, cooperation cannot be sustained in the first place for any value of $(b/c)$.

\subsubsection*{Protection against conditional bribers}

We now consider protection against a more sophisticated bribing strategy---a tax-evading defector who bribes institution members in exchange for a good reputation only if they have been detected as a tax-evader.
We refer to individuals who act on this information as ``conditional'' bribers.
Conditional bribers catch wind of the fact that they are about to be assigned a bad reputation and bribe their way out of it.
Because tax evasion is detected with probability $\D$, the reputation and fitness of ``conditional'' bribers are given by
\begin{equation}
    \begin{split}
        G^\prime_{\text{ALLD}} & = (1 - \D) G_{\text{ALLD}} + \D = \D (1 - G_{\text{ALLD}}) + G_{\text{ALLD}}
        \\
        \Pi^\prime_{\text{ALLD}} & = b G^\prime_{\text{ALLD}} - Q \D \bribesize \salary = b \big[(1 - \D) G_{\text{ALLD}} + \D \big] - \D N \Dbribe \tax ( b - c ) G_{\text{DISC}}.
    \end{split}
\end{equation}
The critical benefit-cost ratio to resist invasion against such a defecting briber is given by
\begin{equation}
    \begin{split}
    \left( \frac{b}{c} \right)^* & = 
    \begin{cases}
        1 + \dfrac{(1 - \D) G_{\text{ALLD}} + \D}{\big[ 1 + \tax (\D N \Dbribe - 1) \big] G_{\text{DISC}} - (1 - \D) G_{\text{ALLD}} - \D } & \text{~when~} N \Dbribe > (N \Dbribe)^*,
        \\
        \infty & \text{~when~} N \Dbribe \leq (N \Dbribe)^*,
    \end{cases}
    \\
    \text{~with~} (N \Dbribe)^* & = \dfrac{1}{\D} + \dfrac{1}{\tax} \dfrac{(1 - G_{\text{ALLD}})}{G_{\text{DISC}}} - \dfrac{1}{\D \tax} \dfrac{G_{\text{DISC}} - G_{\text{ALLD}}}{G_{\text{DISC}}}.
    \label{eq:conditional_resistance_cond}
    \end{split}
\end{equation}
The condition required for finite $(b/c)^*$ here is more complicated, but it hinges again on the relative timescale of strategic updates versus institutional audits, $N \Dbribe$.

When the detection rate $\D$ is very small, conditional bribers are almost never detected as tax evaders---which means they almost never have the opportunity to pay for a good reputation.
They are assessed entirely on the basis of their behavior in pairwise game-play, meaning that, as defectors, they are likely to end up with a bad reputation, and hence low fitness.
Taxpaying discriminators are thus able to resist being invaded by this strategy provided $\tax$ is small.
When the tax rate $\tax$ is very large, tax-paying discriminators have low fitness.
But the tax rate has a much weaker effect on the fitness of tax-evading, conditional bribers.
This is because $\tax$ enters into their fitness only via the bribe amount $\Dbribe$.
For small values of $\D$, such defectors almost never end up paying the bribe anyway.
As a result, when tax evaders are rarely detected, a higher value of $\tax$ raises the fitness of conditional briber-defectors relative to taxpaying discriminators, thus destabilizing cooperation.

When the detection rate $\D$ is large, by contrast, tax evaders are very likely to be detected.
Conditional bribers are thereby likely to offer a bribe---and to end up with a good reputational profile.
The value of the tax rate becomes a double-edged sword.
On the one hand, a higher value of $\tax$ lowers the fitness of discriminators.
On the other hand, it raises the salary $\salary$ and the bribe amount $\bribesize$.
Which one is more important turns out to depend on the compound parameter $\D N \Dbribe$.
If this product is less than $1$, then a higher tax rate makes it more difficult to sustain cooperation, because the effect on the fitness of discriminators is more important.
Alternatively, if $\D N \Dbribe$. is greater than $1$, then the effect of the tax rate on the difficulty of bribing institution members is more important, meaning a higher value of $\tax$ makes bribing a less appealing strategy compared to paying one's taxes.

The benefit-cost ratio required for stable cooperation against conditional bribers can also be understood by  evaluating the partial derivatives of $\Pi_{\text{DISC}} - \Pi_{\text{ALLD}}$;
\begin{equation}
    \begin{split}
        \partial_{\tax} (\Pi^\prime_{\text{DISC}} - \Pi^\prime_{\text{ALLD}}) & = (b - c) G_{\text{DISC}} (\D N \Dbribe - 1),
        \\
        \partial_{\D} (\Pi^\prime_{\text{DISC}} - \Pi^\prime_{\text{ALLD}}) & = (b - c) G_{\text{DISC}} N \Dbribe \tax - b (1 - G_{\text{ALLD}}),
        \\
        \partial_{N \Dbribe} (\Pi^\prime_{\text{DISC}} - \Pi^\prime_{\text{ALLD}}) & = (b - c) \D G_{\text{DISC}} \tax.
    \end{split}
\end{equation}
The derivative with respect to $\tax$ makes explicit our result that, when $\D N \Dbribe > 1$, increasing $\tax$ makes it easier for discriminators to resist invasion, whereas when $\D N \Dbribe < 1$, the opposite holds.

\begin{figure}
    \centering
    \includegraphics[width=\textwidth]{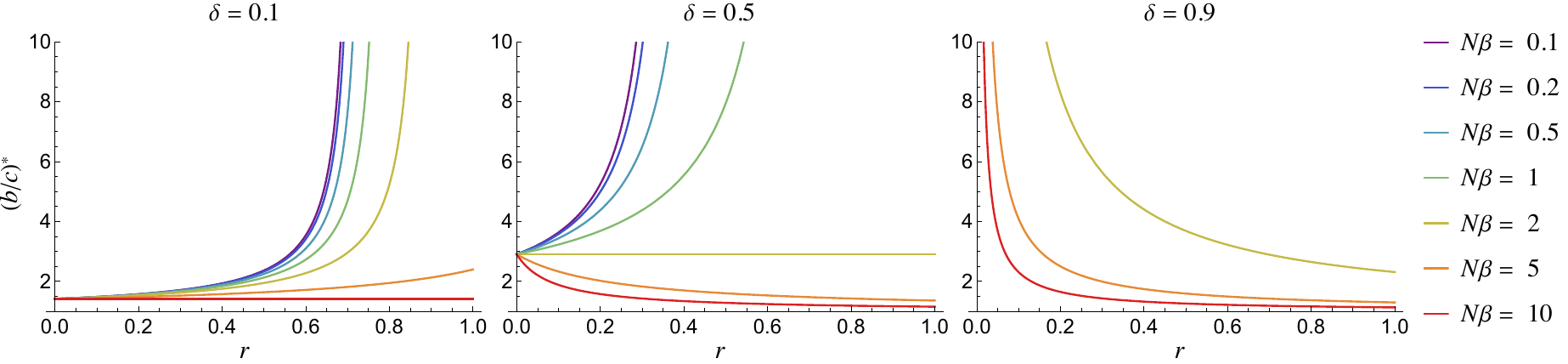}
    \caption{\small{Invisibility of tax-paying discriminators by conditional bribers. Each panel shows the the critical cost-benefit ratio $(b/c)^*$ required to stabilize cooperation, as a function of the tax rate $\tax$, for three different values of the detection probability $\D$.
    The relationship between $(b/c)^*$ and $\tax$ depends on the scaled audit rate parameter $\D N \Dbribe$.
    When $\D N \Dbribe < 1$, the effect of $\tax$ in reducing the fitness of discriminators is more important, and higher taxes destabilize cooperation; when $\D N \Dbribe > 1$, the effect of $\tax$ in increasing the bribe amount is more important, and higher taxes stabilize cooperation.
    (The factor of $\D$ comes from the fact that conditional bribers are detected as tax evaders, and thus given the opportunity to offer a bribe, at rate $\D$).
    }}
    \label{fig:bc_tax_rate_conditional}
\end{figure}

This general result is exemplified by an institution of size $Q = 1$ using \emph{Stern Judging}, in which case
\begin{equation}
    \begin{split}
    \left( \dfrac{b}{c} \right)^* & = 
    \begin{cases}
        1 + \dfrac{2 (1 - \D) \ass (1 - \ass) + \D}{\big[ 1 + \tax (\D N \Dbribe - 1) \big] (1 - \ass) - 2 (1 - \D) \ass (1 - \ass) - \D } & \text{~when~} N \Dbribe > (N \Dbribe)^*,
        \\
        \infty & \text{~when~} N \Dbribe \leq (N \Dbribe)^*,
    \end{cases}
    \\
    \text{~with~} (N \Dbribe^*) & = \dfrac{1}{\D} + \dfrac{1}{\tax} \left( \dfrac{1}{1-\ass} - 2 \ass \right) - \dfrac{1 - 2 \ass}{\D \tax}.
    \label{eq:conditional_resistance_cond}
    \end{split}
\end{equation}
Corresponding results on the stability of cooperation are shown in Figure \ref{fig:bc_tax_rate_conditional}.

\subsection*{Group structured populations}

A critical assumption in our analysis of the maximum tax rate is that, in the absence of an institution, the sole equilibrium is a population consisting entirely of defectors.
However, this may not be the case in real-world populations.
In small societies, for example, other consensus-generating mechanisms, such as gossip \citep{kawakatsu_mechanistic_2024} or empathetic moral evaluation \citep{radzvilavicius_empathy_2019}, may suffice to sustain some level of cooperation.
Such societies will naturally see less value in an institution that guarantees full consensus, because for them, the presence of an institution does less to increase the average payoff.
Moreover, small groups may have their own institutions that suffice for their own purposes but do not scale well to tackling larger problems of coordination, such as the maintenance of resources shared with other societies or the punishment of misdeeds across societal lines.
Scenarios where information is shared locally can be modeled as a group-structured population \citep{nakamura_groupwise_2012,kessinger_evolution_2023}, where individuals agree about reputations with their in-group but not necessarily with their out-group; depending on the social norm, individuals may be much less likely to view their out-group favorably.

We can provide some simple calculations to shed light on what conditions favor the emergence of higher-level institutions in group-structured populations.
Suppose that a society is partitioned into $K$ groups, each comprising a fraction $1/K$ of the population.
Individuals within a group interact at rate $1$, but outgroup interactions occur only with probability $\omega \leq 1$.
We assume that individuals within a group agree about reputations---which can occur as a result of gossip, empathetic perspective-taking, or a local ``institution'' (which, for the sake of simplicity, we assume requires no tax payment)---but that different groups may disagree.
Then the total ingroup cooperation rate, $g^\text{in}$, is likely higher than the outgroup cooperation rate, $g^\text{out}$.
Finally, we allow for different costs ($c^\text{in}$ and $c^\text{out}$) and benefits ($b^\text{in}$ and $b^\text{out}$) for in- versus outgroup interactions.
Without loss of generality, we set $b^\text{in} = b$, $c^\text{in} = c$, $b^\text{out} = \alpha b$, $c^\text{out} = \alpha c$.
We consider $\alpha \geq 1$, meaning that outgroup interactions are more risky but potentially more rewarding.

The average fitness of discriminators and of defectors is now given by
\begin{equation}
    \begin{split}
        \Pi_\text{DISC} & = \frac{(b - c) \left( g_\text{DISC}^\text{in} + \alpha \omega (K-1) g_\text{DISC}^\text{out} \right)}{1 + \omega(K-1)},
        \\
        \Pi_\text{ALLD} & = \frac{b \left( g_\text{ALLD}^\text{in} + \alpha \omega (K-1) g_\text{ALLD}^\text{out}\right) }{1 + \omega(K-1)}.
    \end{split}
\end{equation}
Analytical expressions for the in- and outgroup reputations are derived in Methods.
From these, we can show that discriminators are stable against invasion by defectors provided $b/c$ exceeds the following critical ratio:
\begin{equation}
    \left( \frac{b}{c} \right)^* =
    1 + \frac{g_\text{ALLD}^\text{in} + \alpha \omega (K-1) g_\text{ALLD}^\text{out}}{g_\text{DISC}^\text{in} - g_\text{ALLD}^\text{in} + \alpha \omega (K-1) \left( g_\text{DISC}^\text{out} - g_\text{ALLD}^\text{out} \right)}.
    \label{eq:bc_multigroup}
\end{equation}
(Note that as the number of groups $K$ increases, this ratio approaches infinity; players derive only an infinitesimal portion of their payoff from their in-group, so this is mathematically equivalent to private assessment.)
When the benefit-cost ratio satisfies the condition above, then the group-structured population can sustain cooperation even in the absence of a central institution; in that case, the average fitness in the absence of an institution is not $0$ but rather
\begin{equation}
    \Pi_\text{group-wise} = \frac{(b - c) \left( g_\text{DISC}^\text{in} + \alpha \omega (K-1) g_\text{DISC}^\text{out} \right)}{1 + \omega(K-1)}.
    \label{eq:fitness_group-wise}
\end{equation}

Groups who want to increase their own fitness have several options:
\begin{enumerate}
    \item They may adopt a behavior of always defecting against outgroup members.
    This will raise their fitness in the short run, but if this behavior spreads to other groups via social contagion, the society's average fitness becomes
    $(b - c)g_\text{DISC}^\text{in}/[1 + \omega(K-1)]$; in effect, $g^\text{out}$ will decay to zero.
    \item They may become more insular, i.e., lower their rate of outgroup interactions $\omega$.
    This is guaranteed to increase their own fitness \emph{unless}
        $\alpha > g^\text{in}/g^\text{out}$;
    see Supplementary Information section 5 of \citet{kessinger_evolution_2023}.
    \item They may, in concert with the other groups, create a society-wide institution that broadcasts consensus reputations across all groups.
\end{enumerate}
The last option raises each group's fitness by ensuring consensus about reputations across groups---in effect, causing them to cooperate uniformly at rate $G_\text{DISC}$, rather than having different cooperation rates for in- and out-group interactions.
Note that, when $Q = 1$, the institutional assessment of discriminators and defectors, $G_\text{DISC}$ and $G_\text{ALLD}$, will in fact be equal to $g_\text{DISC}^\text{in}$ and $g_\text{ALLD}^\text{in}$, respectively.
We thus have
\begin{equation}
    \Pi_\text{institution} = \frac{(b - c) g_\text{DISC}^\text{in} \left(1 + \alpha \omega (K-1) \right)}{1 + \omega(K-1)},
\end{equation}
which means that a central institution can substantially increase fitness for all.
In the presence of an institution, groups are likely to enjoy higher fitness by \emph{increasing} their rate of outgroup interaction; the ratio $g^\text{in}/g^\text{out}$ becomes $1$, and so groups will experience evolutionary pressure to engage in rewarding outgroup interactions (i.e., higher $\omega$).
The maximum possible tax rate becomes $\T = \Pi_\text{institution} - \Pi_\text{private}$; the fitness of taxpaying discriminators, who are taxed at rate $\tax$, becomes
\begin{equation}
    \begin{split}
        \Pi_\text{taxpayer} & = (1 - \tax) \Pi_\text{institution} + \tax \Pi_\text{group-wise}
        \\
        & = \frac{(b - c) \left(g^\text{in} + \alpha \omega (K-1) [g^\text{in} + \tax (g^\text{out} - g^\text{in})] \right)}{1 + \omega(K-1)}.
    \label{eq:multigroup_tax}
    \end{split}
\end{equation}
This fitness depends on $\omega$ and $K$ only through the product $\omega (K-1)$; dividing this product by $1 + \omega (K-1)$ yields the portion of an individual's fitness that is due to out-group interactions.

\begin{figure}
    \centering
    \includegraphics[width=\textwidth]{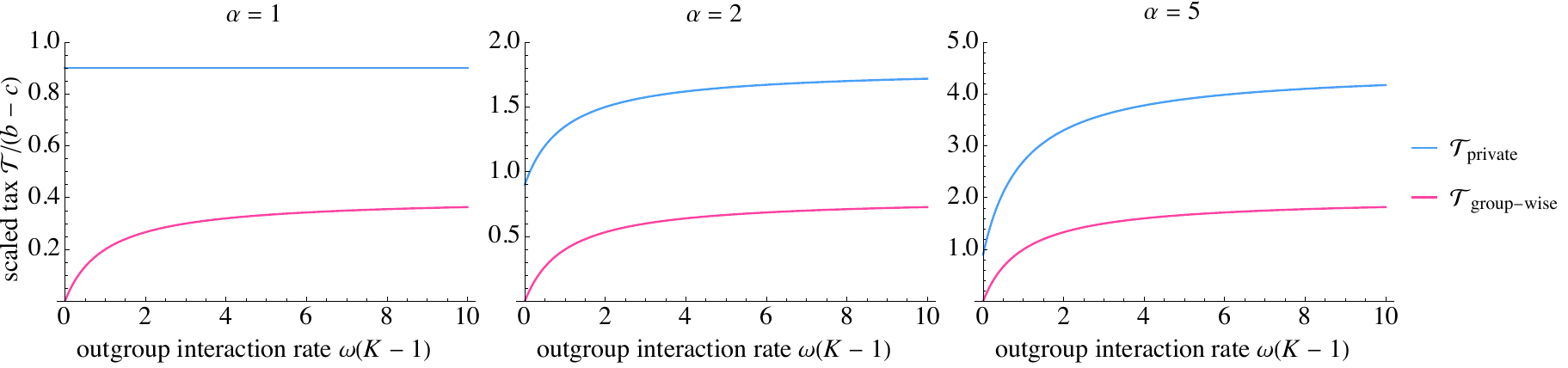}
    \caption{\small{Comparison of the maximal tax rate $\T$, in units of $b - c$, for different information sharing models under Stern Judging.
    If a population cannot sustain cooperation in the absence of an institution––for example, because players rely on private assessment, or because the $b/c$ ratio is not high enough to facilitate cooperation under group-wise information sharing (Equation \ref{eq:bc_multigroup})––each player's expected fitness in the absence of an institution is $0$, and the institution can collect the full sum $\T_\text{ALLD} = \Pi_\text{institution} - \Pi_\text{ALLD}$.
    If group-wise information sharing occurs and is sufficient to sustain some level of cooperation, then the institution can collect only the more modest sum $\T_\text{group-wise} = \Pi_\text{institution} - \Pi_\text{group-wise}$.
    Along the $x$-axis, we vary the compound parameter $\omega (K-1)$, which captures how much of an individual's fitness is determined by out-group interactions; as this fraction increases, the gulf between $\Pi_\text{institution}$ and $\Pi_\text{group-wise}$ increases.
    Different plots correspond to different values of the outgroup interaction premium $\alpha$; in each plot, we set the assessment error $\ass = 0.1$.}}
    \label{fig:tax_comparison}
\end{figure}

Figure \ref{fig:tax_comparison} illustrates the incentives of a population for a centralized institutional broadcast versus group-wise information, as a function of the effective number of groups, $\omega (K-1)$, and the payoff premium associated with out-group interactions, $\alpha$.
The maximum rational tax rate for a central institution is always lower for a group-structured population---because of the possibility for in-group cooperation---than for a population that would defect without any institution.
In other words, group-structured populations that benefit from in-group consensus have less to gain from establishing a centralized institution.
This fundamental result holds irrespective of the rate of outgroup interactions $\omega$ and the relative costs and benefits of out-group interactions, $\alpha$.
We address the fate of tax evaders under such an institutional scheme in the Supplementary Information.

Finally, in the Supplementary Information we consider an additional, more sinister method for generating consensus in a group-structured population: modifying one's social norm so that the out-group is always considered ``bad'' \citep{kessinger_evolution_2023}.
Such a social norm can easily generate some measure of consensus, so a (partially) cooperative equilibrium may exist even in the absence of an institution; it thus reduces the amount a society would be willing to pay to sustain an institution and, in turn, makes the institution more vulnerable to corruption.

\section*{Discussion}

Cooperation is stable only when its benefits accrue disproportionately to those who cooperate.
Under indirect reciprocity, this occurs when reputations are matters of public agreement.
Discriminators who condition their behavior on reputations will reward each other when they agree about who is deserving and who is not.
Thus, consensus is a necessary catalyst for cooperation.
Instead of assuming by fiat that public information is freely available, as is common in prior studies, here we have studied how to maintain an institution that broadcasts information, thereby fomenting the social consensus required for cooperation.

Such an institution can be maintained only because individuals have a personal incentive to support it via taxation---since its absence would lead to universal defection and no payoff for anybody.
We have calculated the maximum tax that rational individuals will contribute to support such an institution.
This rational, ground-up mechanism solves the second-order free-rider problem \citep{yamagishi_provision_1986}: punishing bad behavior is now cost-free, because punishment is meted out by players themselves when they defect against someone with a bad reputation.
We have shown that such an institution can be stable against not only defectors, but also other bad actors---those who shirk their tax obligations, as well as those who seek to bribe institution members in return for a good reputation.
Sufficient institutional effort towards detection allows such individuals to be punished (again, without cost) simply by assigning them a bad reputation.

Human societies are, of course, not uniform and well-mixed, as is often assumed in mathematical models; they may exhibit significant structure and hierarchy, which can impede, or facilitate, cooperation.
Our analysis of group-structured populations allows us to probe these effects.
Partitioning a society into groups, each of which can share information locally, has both positive and negative effects on the maintenance of cooperation.
On the one hand, groups may engage in a high level of internal cooperation, as they exchange information locally.
But at a broader scale, cooperation between groups may be difficult to sustain---even when there is the potential for greater rewards for cooperation across groups.
Local, group-level mechanisms for sustaining cooperation can thus act as a double-edged sword.
Because individuals benefit less from maintaining a centralized institution (compared to the case of no group structure), they will forgo less of their wealth to sustain it; the institution will therefore be more vulnerable to corruption.

Our analysis is certainly not without limitations. Aside from our treatment of group-structured populations, we have otherwise assumed that individuals are essentially interchangeable.
Individuals accrue payoffs only via pairwise game play, and so no individual has an exogenous advantage over any other, for example via greater access to wealth, opportunity, social capital, or information.
And yet, inequality is a pervasive feature of real human societies.
A robust result from the political science literature is that wealth inequality is strongly correlated with a country's propensity toward corrupt governance.
One common interpretation is that corruption facilitates wealth inequality.
However, the arrow of causation might point the opposite direction \citep{policardo_corruption_2018}.
Our analysis favors the latter; the wealthy have stronger incentives to avoid contributing their duly appointed share---and greater access to resources that can help them evade this obligation.
Our analysis could be extended to accommodate variation in underlying wealth, to quantify what extent of inequality is still compatible with a robust institution of reputation judgment.
More generally, relaxing the assumption of exchangeable individuals will be critical in future work on institution-mediated cooperation, both for incorporating more realistic political and economic factors and for considering goals for social justice.

Our analysis of structured populations is also overly simplistic, limited to the effects of group-wise structure alone.
The role of explicit network structure in mediating cooperation via indirect reciprocity remains an area for future research.
Still, our analysis of group-structured populations provides some intuition about how network structure might facilitate cooperation.
When information spreads only locally, a high rate of local cooperation can ensue.
However, to ensure fair play between distant individuals, an institution becomes an attractive option---even more so if interactions among distant individuals are especially fruitful.

We have only begun to scratch the surface when it comes to understanding the interplay between different modes of generating moral consensus.
Gossip provides a ready alternative to institutions insofar as it can ensure agreement about reputations.
However, it must occur quickly enough to compensate for the disagreement induced by independent observations and errors \citep{kawakatsu_mechanistic_2024}.
Depending on the size and structure of a population, this can be a tall order.
On large or sparsely connected networks, for example, information may spread slowly; if interactions occur even between distant individuals on a network, information flow may be too viscous to secure a high rate of cooperation.

Our study extends evolutionary game theory, but it also interacts with a broader literature on the role of governments in society \citep{bowles_cooperative_2011,ostrom_understanding_2005,weber_economy_1978}.
Governments are often seen as having a monopoly on the legitimate use of violence \citep{weber_politik_1926}, which they exercise (in principle) to collect tax revenue and deter other forms of violence.
But institutions in general need not have such a monopoly, or even the ability to use violence at all, in order to be useful.
We focus on a more fundamental role of institutions: generating and maintaining consensus about social reputations in a population.
Socially mediated punishment likely represents an early step in the evolution of human justice; individuals raise their payoffs in pairwise game play by outsourcing their judgment of rightness and wrongness to an central institution.
In turn, institutions that lack the ability to mete out justice to wrongdoers themselves outsource this responsibility to society as a whole.
Phenomena such as outlawry and bounty hunting exemplify the role of institutions as issuers of moral rulings; when an institution cannot back up these rulings with force, that burden falls to the institution's adherents.

On the global stage, as well, international organizations frequently play a role in mediating conflict and providing reputation consensus, even though they lack a direct method to enforce their rulings.
The International Criminal Court, for example, may issue arrest warrants for, and prosecute, individuals accused of crimes against humanity; the International Court of Justice likewise adjudicates disputes between countries.
Neither entity has a standing army or anything resembling a monopoly on violence.
Rather, they rely on the cooperation of their signatory states to enforce their rulings (the ICJ, in particular, relies on the members of the United Nation's Security Council for enforcement)---and so they function much like the institutions we have analyzed here.
This type of institution is not solely a modern phenomenon.
In the pre-modern world, centralized religious authorities could punish by broadcasting judgments––for example, rulings on moral issues––rather than by directly exercising force.
The medieval Catholic Church, for example, gave rise to the ``Pax Dei'' (Peace of God) and ``Treuga Dei'' (Truce of God) movements, which limited the scale of medieval warfare and offered protections for non-combatants (naturally, this protection did not extend to non-believers).
The threat of excommunication was used to compel conformity with church doctrine; even kings and emperors could be expelled from the church, with serious political consequences.
Spiritual punishment may have been the domain of the church, but temporal punishment was the responsibility of the faithful.
All of these examples illustrate that ``information authorities'' supported by compulsory contribution to the public good, of the type we have studied here, are feasible and indeed realistic forms of governance.

The choice between ``local'' mechanisms for information consensus, such as gossip or empathetic perspective taking, and ``global'' mechanisms, like formal institutions, may depend on the nature of the social interactions at play or the size and structure of the population.
In small-scale or low-cost interactions, individuals may rely on informal channels such as word-of-mouth for information exchange.
This localized approach is rooted in trust within close-knit communities.
In high-stakes or riskier transactions, publicly accessible information becomes imperative.
This may manifest through formal institutions that provide structured data, such as credit scores or aggregated customer ratings, enabling individuals to make informed decisions.
This perspective aligns with insights from the economics literature that emphasizes the role of information and trust in shaping interactions \citep{ostrom_governing_1990,granovetter_strength_1973,akerlof_market_1970,north_institutions_1990,coleman_foundations_1990}. 

The transition from local to global mechanisms underscores the dynamic interplay between the scale of economic transactions and the mechanisms employed for information dissemination and trust-building.
We may even envision tension as smaller societies, content with their own institutions, must reckon with the need for larger institutions capable of handling situations of inter-group conflict;
norms enforced by local community consensus may gradually crystallize into formalized mechanisms of governance, which may in turn give way to legal codes and bureaucracies managed by distant authorities.
This tension between different ``levels'' of institutions in group-structured populations is an intriguing topic for future research.

\bigskip

\begin{small}
\section*{Methods}

We consider indirect reciprocity mediated by reputations in a large population of size $N$.
Every round, each individual plays a donation game with every other individual twice, acting once as a donor and once as a recipient.
The donor may either cooperate, paying a cost $c > 0$ to convey a benefit $b > c$ to their interaction partner, or defect, paying no cost and conveying no benefit.
The action of a donor is determined by their behavioral strategy.
We consider two major classes of behavioral strategy: defect (denoted ALLD or $Y$) and discriminate (denoted DISC or $Z$).
Defectors always defect; discriminators cooperate with those who have good reputations and defect with those who have bad reputations.

We then consider two methods for updating reputations, either \emph{private} or \emph{public} assessment.
Under private assessment, the reputation of each individual is updates as follows.
Each observer selects a random interaction in which the focal individual acted as a donor.
The donor is assessed as good or bad depending on their action toward the recipient, the recipient's reputation, and a rule called a \emph{social norm}, which maps this information on to the observer's new view of the donor.
A social norm is characterized by four numbers $P_{RA}$, the probability that the donor is assigned a good reputation after performing action $A \in \{ \text{C}, \text{D} \}$ against a donor who is viewed as $R \in \{ \text{G}, \text{B}\}$ (good or bad).
All values of $P_{RA}$ in our model are either $\ass$ or $1 - \ass$; these correspond to either assigning a bad reputation or a good reputation, but committing an error in assignment (i.e., assigning the wrong reputation) with probability $\ass$.

When assessment is private, each individual has an idiosyncratic, personal view of the rest of the population; in effect, each player's view of another player is an independent, random Bernoulli draw.
Let $f_s$ be the frequency of strategy $s$ and $G_s$ the probability that a strategy $s$ player is viewed as good by an arbitrary observer.
In the large-$N$, ``mean-field'' limit,
\begin{equation}
    \begin{split}
        G & = \sum_s f_s G_s
    \end{split}
\end{equation}
is the fraction of the population viewed as good by an arbitrary observer, and
\begin{equation}
    \begin{split}
        \gamma & = \sum_s f_s G_s^2
    \end{split}
\end{equation}
is the probability that an arbitrary observer and donor agree that a given recipient is good.
We thus have
\begin{equation}
    \begin{split}
        G_{\text{ALLD}} & = g \pgd + (1 - G) \pbd
        \\
        & = g (\pgd - \pbd) + \pbd,
        \\
        G_{\text{DISC}} & = \gamma \pgc + (G - \gamma) (\pgd + \pbc) + (1 - 2G + \gamma)
        \\
        & = \gamma (\pgc - \pgd - \pbc + \pbd) + G (\pgd + \pbc - 2 \pbd) + \pbd.
    \end{split}
    \label{eq:reps_private}
\end{equation}

We next consider public assessment.
After each round of population-wide pairwise game play, the reputation of each individual is updated via broadcast by an \emph{institution}.
An institution consists of $Q$ observers and is characterized by an assessment threshold $q$.
Observers (institution members) pass judgment on an individual just as they do in the private assessment case, with the exception that they rely not on their personal view of the recipient but rather the publicly broadcast institutional view.
Once each observer has passed judgment on the focal individual, these assessments are aggregated into a single reputation as follows: if at least $q$ institution members have assessed an individual as good, their reputation is broadcast to the entire population as good; otherwise, it is broadcast as bad.
The probability that the institution assigns an individual following strategy $s$ a good reputation is thus given by
\begin{equation}
    \begin{split}
        G_s & = \sum_{k = q}^Q {Q \choose k} g_s^k (1 - g_s)^{Q - k}, \text{with}
        \\
        g_{\text{DISC}} & = \pgc G + \pbd (1 - G),
        \\
        g_{\text{ALLD}} & = \pgd G + \pbd (1 - G),
        \\
        G & = \sum_s f_s G_s.
    \end{split}
\end{equation}

Here, $g_s$ is the probability that an individual following strategy $s$ is assessed as good by an arbitrary institution member.
The fitness (total average payoff) of each strategic type is
\begin{equation}
    \begin{split}
        \Pi_{\text{DISC}} & = b f_{\text{DISC}} G_{\text{DISC}} - c G,
        \\
        \Pi_{\text{ALLD}} & = b f_{\text{DISC}} G_{\text{ALLD}}.
    \end{split}
\end{equation}

Finally, after reputations are updated for the entire population, a single individual resolves to update their behavioral strategy.
They do so by choosing a random member of the population and comparing their payoffs.
If the focal individual follows strategy $s$ and their partner follows strategy $s^\prime$, strategy copying occurs with probability given by the Fermi function
\begin{equation}
    \phi(\Pi_s, \Pi_{s^\prime}) = \frac{1}{1 + \exp[ \selstr (\Pi_s - \Pi_{s^\prime})]}.
\end{equation}
In the limit of weak selection (small $\selstr$), it can be shown that this yields a replicator equation for the dynamics of strategic types \citep{kessinger_evolution_2023}; however, since we are primarily concerned with whether or not rare mutants can invade, we restrict our analysis to whether a mutant (with strategy $s^\prime$) is fitter than a population of residents (with strategy $s$).
In general, discriminators resist invasion by defectors provided
\begin{equation}
    \frac{b}{c} > 1 + \frac{G_{\text{ALLD}}}{G_{\text{DISC}} - G_{\text{ALLD}}}.
    \label{eq:bccond_methods}
\end{equation}

In the absence of an institution, individuals would not be guaranteed to agree on reputations, so they would have to rely on their own private assessments of each others' moral standing.
Under the social norm \emph{Stern Judging}, private assessment cannot sustain cooperation; individuals have their own idiosyncratic views of the rest of the population.
The solution to Equation \ref{eq:reps_private} is $G_s = 1/2$, irrespective of the strategy $s$ and the composition of the population.
Accordingly, Equation \ref{eq:bccond_methods} blows up.
The sole stable equilibrium is a population of defectors, meaning each individual accrues a payoff of zero:
\begin{equation}
    \Pi_\text{private} = 0.
\end{equation}
In the presence of an institution, the payoff can be much higher \citep{radzvilavicius_adherence_2021}; institutions foment agreement about reputations, meaning discriminators may behave in ways that cause them to be viewed as good by other discriminators.
Thus, there is a stable equilibrium consisting entirely of discriminators:
\begin{equation}
    \Pi_\text{institution} = (b - c) G_{\text{DISC}}|_{f_{\text{DISC}} = 1}.
\end{equation}

Institution members may well demand payment for the value they create for the population by broadcasting consensus reputations.
This value is precisely equal to the difference between the individual payoff when there is no institution and when there is an institution.
Provided the tax amount is smaller than this difference, the population will be willing to pay it.
The maximum per capita tax is thus
\begin{equation}
    \mathcal{T} = \Pi_\text{institution} - \Pi_\text{private} = (b - c) G_{\text{DISC}}|_{f_{\text{DISC}} = 1}.
\end{equation}
Suppose the tax levied is some fraction $\tax$ of this maximum amount.
The fitness of discriminators who pay this tax is then
\begin{equation}
    \Pi_{\text{DISC}}^\prime = (1 - \tax) (b - c) G_{\text{DISC}}.
\end{equation}
If tax revenue is then disbursed equally among institution members, the per-round salary earned by each institution member is
\begin{equation}
    \salary = N \tax (b - c) G_{\text{DISC}}/Q.
\end{equation}

We now posit that defectors not only defect in every interaction but also refuse to pay the tax; with probability $\D$, their tax evasion is discovered, and they are automatically assigned a bad reputation for that round.
The fitness of such a defector, in a population otherwise consisting entirely of discriminators, is
\begin{equation}
    \begin{split}
        \Pi_{\text{ALLD}}^\prime & = b G_{\text{ALLD}}^\prime, \text{with}
        \\
        G_{\text{ALLD}}^\prime & = (1 - \D) G_{\text{ALLD}}.
    \end{split}
\end{equation}

Finally, we consider the possibility of institutional corruption.
With probability $\Dbribe$, a given institution member is audited for possible corruption; if they are found to be corrupt, they are removed from the institution and must forego their salary.
Corruption in our model takes the form of bribes paid by defectors in exchange for a good reputation.
In order for an institution member to accept a bribe, their expected gain from taking the bribe must outweigh their expected loss.
This means the bribe amount must be greater than the product of the salary and the audit risk.
\begin{equation}
    \bribesize > \Dbribe \salary = N \Dbribe \tax (b-c) G_{\text{DISC}} / Q.
\end{equation}

As a further consideration, we analyze the behavior of an institution when the population \emph{is} capable of sustaining some level of cooperation.
We focus on the scenario of a population partitioned into $K$ groups, such that each group is capable of maintaining consensus about reputations via, e.g., rapid gossip.
Individuals in each group interact with their in-group at rate $1$ and their out-group at rate $1 \geq \omega \geq 0$.
The reputations of defectors and discriminators in this scenario are given by
\begin{equation}
    \begin{split}
        G_\text{ALLD}^\text{in} & = \frac{G^\text{in} + \omega (K-1) G^\text{out}}{1 + \omega (K-1) } (\pgd - \pbd) + \pbd,
        \\
        G_\text{ALLD}^\text{out} & = \frac{\omega G^\text{in} + [1 + \omega (K-2)] G^\text{out}}{1 + \omega (K-1) } (\pgd - \pbd) + \pbd,
        \\
        G_\text{DISC}^\text{in} & = \frac{G^\text{in} + \omega (K-1) G^\text{out}}{1 + \omega (K-1) } (\pgc - \pbd) + \pbd,
        \\
        G_\text{DISC}^\text{out} & = \frac{[1 + \omega] G^\text{in} G^\text{out} + \omega [K - 2](G^\text{out})^2}{1 + \omega (K-1) }(\pgc - \pgd - \pbc + \pbd)
        \\
        & \ \ \ + \frac{ \omega G^\text{in} + [1 + \omega (K-2) ] G^\text{out}} {1 + \omega (K-1) }(\pgd-\pbd)
        \\
        & \ \ \ + \frac{ G^\text{in} + \omega(K-1) G^\text{out}} {1 + \omega (K-1) } (\pbc - \pbd) + \pbd;
    \end{split}
\end{equation}
see section $5$ of the supplement of \citet{kessinger_evolution_2023}.
Under Stern Judging, the solutions are
\begin{equation}
    \begin{split}
    g_\text{DISC}^\text{in} = 1 - \ass, & \quad g_\text{ALLD}^\text{in} = \frac{4 \ass (1 - \ass) + \omega(K-1)}{2[1+\omega(K-1)]},
    \\
    g_\text{DISC}^\text{out}= \frac{1}{2}, & \quad g_\text{ALLD}^\text{out} = \frac{1 + \omega [4\ass(1 - \ass) + K - 2]}{2[1 + \omega(K-1)]}.
    \end{split}
\end{equation}
Note that, as the product $\omega (K-1)$ increases, both $g_\text{ALLD}^\text{in}$ and $g_\text{ALLD}^\text{out}$ approach $1/2$, consistent with private assessment.

When groups can sustain cooperation even in the absence of an institution, adding an institution has a reduced effect on their fitness compared to the case of private assessment.
Under Stern Judging, we have $\Pi_\text{private} = 0$, $\Pi_\text{institution} = 1 - \ass$, and $\Pi_\text{group-wise}$ given by equation $\ref{eq:fitness_group-wise}$, with $g_\text{DISC}^\text{in} = 1 - \ass$, $g_\text{DISC}^\text{out} = 1/2$.
Imposing an institution increases the population's fitness by
\begin{equation}
    \Pi_\text{institution} - \Pi_\text{group-wise} = \frac{(b - c) \alpha \omega (K-1)(g^\text{in} - g^\text{out})}{1 + \omega(K-1)},
\end{equation}
which is precisely the maximum tax rate $\mathcal{T}$ the population will be willing to pay in this scenario.

\bibliographystyle{abbrvnat}
\bibliography{references}

\end{small}
\end{document}


\maketitle

\section{Private, public, and group-wise reputation dynamics}

We begin by outlining the reputation dynamics under the three major classes of reputation assignment: private assessment, public assessment, and group-wise reputation assessment.

As in the main text, we define
\begin{equation}
    \begin{split}
        G & = \sum_s f_s G_s,
        \\
        \gamma & = \sum_s f_s G_s^2,
    \end{split}
\end{equation}
with $s \in \{ \text{ALLC}, \text{ALLD}, \text{DISC} \}$.
(The ALLC strategy becomes important under norms other than Stern Judging.)

\subsection{Private assessment}

The reputations of each strategic type may be obtained as follows:
\begin{enumerate}
    \item \textbf{ALLC}. A cooperator interacts with a randomly chosen recipient:
    \begin{enumerate}
        \item With probability $G$, the observer views the recipient as good.
        The cooperator cooperates; the observer assigns them a good reputation with probability $\pgc$.
        \item With probability $1 - G$, the observer views the recipient as good.
        The cooperator cooperates; the observer assigns them a good reputation with probability $\pbc$.
    \end{enumerate}
    \item \textbf{ALLD}. A defector interacts with a randomly chosen recipient:
    \begin{enumerate}
        \item With probability $G$, the observer views the recipient as good.
        The defector defects; the observer assigns them a good reputation with probability $\pgd$.
        \item With probability $1 - G$, the observer views the recipient as good.
        The cooperator cooperates; the observer assigns them a good reputation with probability $\pbd$.
    \end{enumerate}
    \item \textbf{DISC}. A discriminator interacts with a randomly chosen recipient.
    With probability $f_s$, the recipient follows strategy $s$:
    \begin{enumerate}
        \item With probability $G_s$, the observer views the recipient as good.
        With probability $G_s$, the discriminator views the recipient as good and cooperates.
        The observer assigns them a good reputation with probability $\pgc$.
        \item With probability $G_s$, the observer views the recipient as good.
        With probability $1 - G_s$, the discriminator views the recipient as bad and defects.
        The observer assigns them a good reputation with probability $\pgd$.
        \item With probability $1 - G_s$, the observer views the recipient as good.
        With probability $G_s$, the discriminator views the recipient as good and cooperates.
        The observer assigns them a good reputation with probability $\pbc$.
        \item With probability $1 - G_s$, the observer views the recipient as good.
        With probability $1 - G_s$, the discriminator views the recipient as bad and defects.
        The observer assigns them a good reputation with probability $\pbd$.
    \end{enumerate}
\end{enumerate}
Summing over all strategic types yields $\sum_s f_s G_s^2 = \gamma$.
Thus, under private assessment, we have
\begin{equation}
    \begin{split}
        G_\text{ALLC} & = G \pgc + (1 - G) \pbc
        \\
        & = G (\pgc - \pbc) + \pbc,
        \\
        G_{\text{ALLD}} & = G \pgd + (1 - G) \pbd
        \\
        & = G (\pgd - \pbd) + \pbd,
        \\
        G_{\text{DISC}} & = \gamma \pgc + (G - \gamma) (\pgd + \pbc) + (1 - 2G + \gamma)
        \\
        & = \gamma (\pgc - \pgd - \pbc + \pbd) + G (\pgd + \pbc - 2 \pbd) + \pbd.
    \end{split}
    \label{eq:reps_private}
\end{equation}

\subsection{Public (institutional) assessment.}
The reputations of each strategic type under institutional public assessment may be obtained as follows.
As in the main text, $Q$ institution members assess players independently; if at least $q$ of them regard a player as good, their reputation is broadcast to the entire population as good.
Let $g_s$ be the probability that an arbitrary institution member views a strategy $s$ player as good.
\begin{enumerate}
    \item \textbf{ALLC}. A cooperator interacts with a randomly chosen recipient:
    \begin{enumerate}
        \item With probability $G$, the institution member views the recipient as good.
        The cooperator cooperates; the institution member assigns them a good reputation with probability $\pgc$.
        \item With probability $1 - G$, the institution member views the recipient as good.
        The cooperator cooperates; the institution member assigns them a good reputation with probability $\pbc$.
    \end{enumerate}
    \item \textbf{ALLD}. A defector interacts with a randomly chosen recipient:
    \begin{enumerate}
        \item With probability $G$, the institution member views the recipient as good.
        The defector defects; the institution member assigns them a good reputation with probability $\pgd$.
        \item With probability $1 - G$, the institution member views the recipient as good.
        The cooperator cooperates; the institution member assigns them a good reputation with probability $\pbd$.
    \end{enumerate}
    \item \textbf{DISC}. A discriminator interacts with a randomly chosen recipient.
    With probability $f_s$, the recipient follows strategy $s$:
    \begin{enumerate}
        \item With probability $G$, the institution member views the recipient as good.
        The discriminator cooperates; the institution member assigns them a good reputation with probability $\pgc$.
        \item With probability $1 - G$, the institution member views the recipient as good.
        The discriminator defects; the institution member assigns them a good reputation with probability $\pbd$.
    \end{enumerate}
\end{enumerate}
Thus, under institutional assessment, we have
\begin{equation}
    \begin{split}
        g_\text{ALLC} & = G \pgc + (1 - G) \pbc
        \\
        & = G (\pgc - \pbc) + \pbc,
        \\
        g_{\text{ALLD}} & = G \pgd + (1 - G) \pbd
        \\
        & = G (\pgd - \pbd) + \pbd,
        \\
        g_{\text{DISC}} & = G \pgc + (1 - G) \pbd
        \\
        & = G (\pgc - \pbd) + \pbd, \text{with}
        \\
        G_s & = \sum_{k = q}^Q {Q \choose k} g_s^k (1 - g_s)^{Q - k}
    \end{split}
    \label{eq:reps_institutional}
\end{equation}
and $G = \sum_s f_s G_s$ as usual.
Note that if $Q = q = 1$, institutional public assessment is mathematically identical to the standard model of public assessment in indirect reciprocity, in which the entire population is compelled to agree on everyone else's reputation--for example, by rapid gossip or empathetic perspective-taking.

\subsection{Group-wise assessment}

Finally, we consider a model in which information about reputations flows freely within group lines––for example, because of rapid within-group gossip, empathetic perspective-taking within group lines, or group-wise ``institutions'' that do not broadcast information outside of their groups.
Every player in a group agrees about everyone else's reputations, but different groups may disagree.
For simplicity, we assume there are $K$ groups of equal size $1/K$; interactions between members of the same group occur with probability $1$, but potential dyadic interactions between members of different groups occur only with probability $1 \geq \omega \geq 0$.
Note that the average reputation of a given strategic type is given by
\begin{equation}
    G_s = \frac{G_s^\text{in} + \omega (K-1) G_s^\text{out}}{1 + \omega (K-1)}.
\end{equation}
For simplicity, we consider \emph{only} the case where the strategy frequencies $f_s$ are the same in all groups; a full treatment for arbitrary strategic frequencies may be found in Section 5 of the supplement of \citet{kessinger_evolution_2023}.
\begin{enumerate}
    \item \textbf{ALLC}. A cooperator interacts with a randomly chosen recipient:
    \begin{enumerate}
        \item \textbf{In-group reputation}. If the observer is in the same group as the donor, then:
        \begin{enumerate}
            \item With probability $G$, the observer views the recipient as good.
            The cooperator cooperates; the observer assigns them a good reputation with probability $\pgc$.
            \item With probability $1 - G$, the observer views the recipient as bad.
            The cooperator cooperates; the observer assigns them a good reputation with probability $\pbc$.
        \end{enumerate}
        \item \textbf{Out-group reputation}. If the observer is in a different group from the donor, then:
        \begin{enumerate}
            \item With probability $\omega/(1 + \omega (K-1))$, the recipient is in the observer's group.
            \begin{enumerate}
                \item With probability $G^\text{in}$, the observer views the recipient as good.
                The cooperator cooperates; the observer assigns them a good reputation with probability $\pgc$.
                \item With probability $1 - G^\text{in}$, the observer views the recipient as bad.
                The cooperator cooperates; the observer assigns them a good reputation with probability $\pbc$.
            \end{enumerate}
            \item With probability $1/(1 + \omega (K-1))$, the recipient is in the donor's group.
            \begin{enumerate}
                \item With probability $G^\text{out}$, the observer views the recipient as good.
                The cooperator cooperates; the observer assigns them a good reputation with probability $\pgc$.
                \item With probability $1 - G^\text{out}$, the observer views the recipient as bad.
                The cooperator cooperates; the observer assigns them a good reputation with probability $\pbc$.
            \end{enumerate}
            \item With probability $\omega(K-2)/(1 + \omega(K-1))$, the recipient is in neither the donor's group nor the observer's group.
            \begin{enumerate}
                \item With probability $G^\text{out}$, the observer views the recipient as good.
                The cooperator cooperates; the observer assigns them a good reputation with probability $\pgc$.
                \item With probability $1 - G^\text{out}$, the observer views the recipient as bad.
                The cooperator cooperates; the observer assigns them a good reputation with probability $\pbc$.
            \end{enumerate}
        \end{enumerate}
    \end{enumerate}
    Thus
    \begin{equation}
        \begin{split}
        G_\text{ALLC}^\text{in} & = G \pgc + (1 - G) \pbc
        \\
        & = G (\pgc - \pbc) + \pbc,
        \\
        G_\text{ALLC}^\text{out} & = \frac{\omega [G^\text{in} \pgc + (1 - G^\text{in}) \pbc] + [1 + \omega(K-2][G^\text{out} \pgc + (1 - G^\text{out}) \pbc]}{1 + \omega(K-1)}
        \\
        & = \frac{\omega G^\text{in} [1 + \omega(K-2] G^\text{out}}{1 + \omega(K-1)}(\pgc - \pbc) + \pbc.
        \end{split}
    \end{equation}
    \item \textbf{ALLD}. A defector interacts with a randomly chosen recipient:
    \begin{enumerate}
        \item \textbf{In-group reputation}. If the observer is in the same group as the donor, then:
        \begin{enumerate}
            \item With probability $G$, the observer views the recipient as good.
            The defector defects; the observer assigns them a good reputation with probability $\pgd$.
            \item With probability $1 - G$, the observer views the recipient as bad.
            The defector defects; the observer assigns them a good reputation with probability $\pbd$.
        \end{enumerate}
        \item \textbf{Out-group reputation}. If the observer is in a different group from the donor, then:
        \begin{enumerate}
            \item With probability $\omega/(1 + \omega (K-1))$, the recipient is in the observer's group.
            \begin{enumerate}
                \item With probability $G^\text{in}$, the observer views the recipient as good.
                The defector defects; the observer assigns them a good reputation with probability $\pgd$.
                \item With probability $1 - G^\text{in}$, the observer views the recipient as bad.
                The defector defects; the observer assigns them a good reputation with probability $\pbd$.
            \end{enumerate}
            \item With probability $1/(1 + \omega (K-1))$, the recipient is in the donor's group.
            \begin{enumerate}
                \item With probability $G^\text{out}$, the observer views the recipient as good.
                The defector defects; the observer assigns them a good reputation with probability $\pgd$.
                \item With probability $1 - G^\text{out}$, the observer views the recipient as bad.
                The defector defects; the observer assigns them a good reputation with probability $\pbd$.
            \end{enumerate}
            \item With probability $\omega(K-2)/(1 + \omega(K-1))$, the recipient is in neither the donor's group nor the observer's group.
            \begin{enumerate}
                \item With probability $G^\text{out}$, the observer views the recipient as good.
                The defector defects; the observer assigns them a good reputation with probability $\pgd$.
                \item With probability $1 - G^\text{out}$, the observer views the recipient as bad.
                The defector defects; the observer assigns them a good reputation with probability $\pbd$.
            \end{enumerate}
        \end{enumerate}
    \end{enumerate}
    Thus
    \begin{equation}
        \begin{split}
        G_\text{ALLD}^\text{in} & = G \pgd + (1 - G) \pbd
        \\
        & = G (\pgd - \pbd) + \pbd,
        \\
        G_\text{ALLD}^\text{out} & = \frac{\omega [G^\text{in} \pgd + (1 - G^\text{in}) \pbd] + [1 + \omega(K-2][G^\text{out} \pgd + (1 - G^\text{out}) \pbd]}{1 + \omega(K-1)}
        \\
        & = \frac{\omega G^\text{in} [1 + \omega(K-2] G^\text{out}}{1 + \omega(K-1)}(\pgd - \pbd) + \pbd.
        \end{split}
    \end{equation}
    \item \textbf{DISC}. A discriminator interacts with a randomly chosen recipient.
    \begin{enumerate}
        \item \textbf{In-group reputation}. If the observer is in the same group as the donor, then:
        \begin{enumerate}
            \item With probability $G$, the observer views the recipient as good.
            The discriminator cooperates; the observer assigns them a good reputation with probability $\pgc$.
            \item With probability $1 - G$, the observer views the recipient as bad.
            The discriminator defects; the observer assigns them a good reputation with probability $\pbd$.
        \end{enumerate}
        \item \textbf{Out-group reputation}.
        With probability $f_s$, the recipient follows strategy $s$.
        If the observer is in a different group from the discriminator, then:
        \begin{enumerate}
            \item With probability $\omega/(1 + \omega (K-1))$, the recipient is in the observer's group.
            \begin{enumerate}
                \item With probability $G_s^\text{in}$, the observer views the recipient as good.
                With probability $G_s^\text{out}$, the discriminator views the recipient as good.
                The discriminator cooperates; the observer assigns them a good reputation with probability $\pgc$.
                \item With probability $G_s^\text{in}$, the observer views the recipient as good.
                With probability $1 - G_s^\text{out}$, the discriminator views the recipient as bad.
                The discriminator defects; the observer assigns them a good reputation with probability $\pgd$.
                \item With probability $1 - G_s^\text{in}$, the observer views the recipient as bad.
                With probability $G_s^\text{out}$, the discriminator views the recipient as good.
                The discriminator cooperates; the observer assigns them a good reputation with probability $\pbc$.
                \item With probability $1 - G_s^\text{in}$, the observer views the recipient as bad.
                With probability $1 - G_s^\text{out}$, the discriminator views the recipient as bad.
                The discriminator defects; the observer assigns them a good reputation with probability $\pbd$.
            \end{enumerate}
            \item With probability $1/(1 + \omega (K-1))$, the recipient is in the donor's group.
            \begin{enumerate}
                \item With probability $G_s^\text{out}$, the observer views the recipient as good.
                With probability $G_s^\text{in}$, the discriminator views the recipient as good.
                The discriminator cooperates; the observer assigns them a good reputation with probability $\pgc$.
                \item With probability $G_s^\text{out}$, the observer views the recipient as good.
                With probability $1 - G_s^\text{in}$, the discriminator views the recipient as bad.
                The discriminator defects; the observer assigns them a good reputation with probability $\pgd$.
                \item With probability $1 - G_s^\text{out}$, the observer views the recipient as bad.
                With probability $G_s^\text{in}$, the discriminator views the recipient as good.
                The discriminator cooperates; the observer assigns them a good reputation with probability $\pbc$.
                \item With probability $1 - G_s^\text{out}$, the observer views the recipient as bad.
                With probability $1 - G_s^\text{in}$, the discriminator views the recipient as bad.
                The discriminator defects; the observer assigns them a good reputation with probability $\pbd$.
            \end{enumerate}
            \item With probability $\omega(K-2)/(1 + \omega(K-1))$, the recipient is in neither the donor's group nor the observer's group.
            \begin{enumerate}
                \item With probability $G_s^\text{out}$, the observer views the recipient as good.
                With probability $G_s^\text{out}$, the discriminator views the recipient as good.
                The discriminator cooperates; the observer assigns them a good reputation with probability $\pgc$.
                \item With probability $G_s^\text{out}$, the observer views the recipient as good.
                With probability $1 - G_s^\text{out}$, the discriminator views the recipient as bad.
                The discriminator defects; the observer assigns them a good reputation with probability $\pgd$.
                \item With probability $1 - G_s^\text{out}$, the observer views the recipient as bad.
                With probability $G_s^\text{out}$, the discriminator views the recipient as good.
                The discriminator cooperates; the observer assigns them a good reputation with probability $\pbc$.
                \item With probability $1 - G_s^\text{out}$, the observer views the recipient as bad.
                With probability $1 - G_s^\text{out}$, the discriminator views the recipient as bad.
                The discriminator defects; the observer assigns them a good reputation with probability $\pbd$.
            \end{enumerate}
        \end{enumerate}
    \end{enumerate}
    Thus
    \begin{equation}
        \begin{split}
        G_\text{DISC}^\text{in} & = G \pgc + (1 - G) \pbd
        \\
        & = G (\pgc - \pbd) + \pbd,
        \\
        G_\text{DISC}^\text{out} & = \sum_s f_s \Bigg( \frac{\omega [G_s^\text{in} G_s^\text{out} \pgc + G_s^\text{in} (1 - G_s^\text{out}) \pgd + (1 - G_s^\text{in}) G_s^\text{out} \pbc + (1 - G_s^\text{in}) (1 - G_s^\text{out}) \pbd]}{1 + \omega(K-1)}
        \\
        & \ \ \ \ \ + \frac{G_s^\text{out} G_s^\text{in} \pgc + G_s^\text{out} (1 - G_s^\text{in}) \pgd + (1 - G_s^\text{out}) G_s^\text{in} \pbc + (1 - G_s^\text{out}) (1 - G_s^\text{in}) \pbd}{1 + \omega(K-1)}
        \\
        & \ \ \ \ \ + \frac{\omega (K-2) [(G_s^\text{out})^2 \pgc + G_s^\text{out} (1 - G_s^\text{out}) \pgd + (1 - G_s^\text{out}) G_s^\text{out} \pbc + (1 - G_s^\text{out})^2 \pbd]}{1 + \omega(K-1)} \Bigg)
        \\
        & = \sum_s f_s \Bigg( \frac{(1 + \omega) G_s^\text{in} G_s^\text{out} + \omega (K-2) (G_s^\text{out})^2}{1 + \omega(K-1)} (\pgc - \pgd - \pbc + \pbd)
        \\
        & \ \ \ \ \ + \frac{ \omega G_s^\text{in} + [1 + \omega (K-2) ] G_s^\text{out}} {1 + \omega (K-1) }(\pgd - \pbd)
        \\
        & \ \ \ \ \ + \frac{ G_s^\text{in} + \omega(K-1) G_s^\text{out}} {1 + \omega (K-1) } (\pbc - \pbd) + \pbd \Bigg).
        \end{split}
    \end{equation}
\end{enumerate}

\section{Norm-specific results under private assessment}

Fitnesses of each strategic type are given by
\begin{equation}
    \begin{split}
        \Pi_\text{ALLC} & = b (f_\text{ALLC} + f_\text{DISC} G_\text{ALLC} ) - c,
        \\
        \Pi_\text{ALLD} & = b (f_\text{ALLC} + f_\text{DISC} G_\text{ALLD} ),
        \\
        \Pi_\text{DISC} & = b (f_\text{ALLC} + f_\text{DISC} G_\text{DISC} ) - c G.
    \end{split}
    \label{eq:fitnesses}
\end{equation}
Irrespective of the social norm, we find that
\begin{equation}
    f_\text{ALLD} = 1
\end{equation}
is a stable equilibrium under any social norm and any mode of reputation assignment.
However, some norms allow for a nonzero basin of attraction toward an equilibrium that sustains some level of cooperation.
We explore these possibilities here.

\subsection{Arbitrary second-order norms.}

We begin by considering second-order norms with $\pgc = 1 - \ass, \pgd = \ass$ as members of a continuous family of norms, with $\pbc = \pc(1 - 2\ass) + \ass, \pbc = \pd(1 - 2\ass) + \ass$.
Here, $\pc$ represents the probability that cooperating with a player with a bad reputation yields a good reputation (up to errors), and $\pd$ represents the probability that defecting with a player with a bad reputation yields a good reputation (again, up to errors).
Setting $(\pc, \pd) = (0,0), (0,1), (1,0), (1,1)$ yields the norms Shunning, Stern Judging, Scoring, and Simple Standing, respectively.
Equation \ref{eq:reps_private} becomes
\begin{equation}
    \begin{split}
        G_\text{ALLC} & = G (1 - 2 \ass)(1 - \pc) + \pc (1 - 2\ass) + \ass ,
        \\
        G_{\text{ALLD}} & = G (2 \ass - 1)(1 - \pd) + \pd (1 - 2\ass) + \ass ,
        \\
        G_{\text{DISC}} & = \gamma (1 - \pc + \pd)(1 - 2\ass) + G (\pc - 2 \pd)(1 - 2\ass) + \pc(1 - 2\ass) + \ass.
    \end{split}
\end{equation}
The stability of the $f_\text{DISC} = 1$ equilibrium is determined as follows.
If $b/c$ exceeds
\begin{equation}
    \left( \frac{b}{c}\right)^* = \frac{2}{1 + \pc(1 - 2\ass) + \sqrt{(1-\pc)^2 + 4[(1 - \pc)^2 + \pd] \ass + 4[2 + \pc(2 - \pc) - 2 \pd) \ass^2]}},
\end{equation}
then DISC resists invasion by ALLD but \emph{not} by ALLC.
Conversely, if $b/c$ is \emph{less} than $(b/c)^*$, then DISC resists invasion by ALLC but \emph{not} by ALLD.
In the former case, there may be a basin of attraction toward a mixed DISC-ALLC equilibrium, corresponding to
\begin{equation}
    \begin{split}
        f_\text{ALLC}^* & = \frac{b^2(1 - 2\ass)[\pc + (1 - \pc - \pd) \ass] - bc[1 + p(1 - 2\ass)] + c^2}{b(1 - 2\ass)(b [\pc + (1 - \pc - \pd) \ass] - c )},
        \\
        f_\text{DISC}^* & = 1 - f_\text{ALLC}^*.
    \end{split}
\end{equation}
Cooperation then occurs at rate $f_\text{ALLC}^* + f_\text{DISC}^* G$.

\subsection{Stern Judging.}
Under Stern Judging, Equation \ref{eq:reps_private} becomes
\begin{equation}
    \begin{split}
        G_\text{ALLC} & = G (1 - 2 \ass) + \ass,
        \\
        G_{\text{ALLD}} & = G (2 \ass - 1) + 1 - \ass,
        \\
        G_{\text{DISC}} & = 2 \gamma (1 - 2 \ass) - 2 G (1 - 2 \ass) + 1 - \ass.
    \end{split}
\end{equation}
For $1/2 > \ass > 0$, this has only one solution where all the $G$ fall within $[0, 1]$, namely
\begin{equation}
    G_s = \frac{1}{2} \text{ for all $s$},
\end{equation}
irrespective of the strategy frequencies $f_s$.
As a result, Equation \ref{eq:fitnesses} becomes
\begin{equation}
    \begin{split}
        \Pi_\text{ALLC} & = b (f_\text{ALLC} + \frac{1}{2} f_\text{DISC} ) - c,
        \\
        \Pi_\text{ALLD} & = b (f_\text{ALLC} + \frac{1}{2} f_\text{DISC}),
        \\
        \Pi_\text{DISC} & = b (f_\text{ALLC} + \frac{1}{2} f_\text{DISC} ) - \frac{c}{2}.
    \end{split}
\end{equation}
The relation $\Pi_\text{ALLD} > \Pi_\text{DISC} > \Pi_\text{ALLC}$ always obtains, so $f_\text{ALLD} = 1$ is the sole stable equilibrium.

\subsection{Shunning.} Under Shunning, Equation \ref{eq:reps_private} becomes
\begin{equation}
    \begin{split}
        G_\text{ALLC} & = G (1 - 2 \ass) + \ass,
        \\
        G_{\text{ALLD}} & = \ass,
        \\
        G_{\text{DISC}} & = \gamma (1 - 3 \ass) + \ass.
    \end{split}
\end{equation}
The general solution is complicated, but in a population of discriminators, we have
\begin{equation}
    \begin{split}
        G_\text{ALLC} & = \frac{1 + 2\ass - \sqrt{1 - 4\ass + 8 \ass^2}}{2},
        \\
        G_{\text{ALLD}} & = \ass,
        \\
        G_{\text{DISC}} & = \frac{1 - \sqrt{1 - 4\ass + 8 \ass^2}}{2(1 - 2\ass)}.
    \end{split}
\end{equation}
A population of discriminators is stable against invasion by defectors provided $b/c$ exceeds
\begin{equation}
    \left( \frac{b}{c} \right)^* = \frac{2}{1 - \sqrt{1 - 4\ass + 8 \ass^2}}.
\end{equation}
At exactly this cutoff, however, it becomes vulnerable to invasion by \emph{cooperators}.
There is a basin of attraction toward a mixed equilibrium of discriminators and cooperators at
\begin{equation}
    f_\text{ALLC} = \frac{b c - c^2 - b^2 \ass (1 - 2\ass)}{[bc - (b^2 + c^2) \ass](1 - 2 \ass)},
\end{equation}
at which the rate of cooperation is
\begin{equation}
    f_\text{ALLC} + f_\text{DISC} G = \frac{b c - c^2 - b^2 \ass (1 - 2\ass)}{[bc - (b^2 + c^2) \ass](1 - 2 \ass)} + \Big(1 - \frac{b c - c^2 - b^2 \ass (1 - 2\ass)}{[bc - (b^2 + c^2) \ass](1 - 2 \ass)} \Big) \Big( \frac{c - b \ass}{c - 2 b \ass} \Big).
\end{equation}

\subsection{Simple Standing.} Under Simple Standing, Equation \ref{eq:reps_private} becomes
\begin{equation}
    \begin{split}
        G_\text{ALLC} & = 1 - \ass,
        \\
        G_{\text{ALLD}} & = G (2 \ass - 1) + 1 - \ass
        \\
        G_{\text{DISC}} & = \gamma (1 - 2 \ass) + G(2 \ass - 1) + 1 - \ass.
    \end{split}
\end{equation}
In a population of discriminators, we have
\begin{equation}
    \begin{split}
        G_\text{ALLC} & = 1 - \ass,
        \\
        G_{\text{ALLD}} & = \sqrt{\ass(1 - \ass)},
        \\
        G_{\text{DISC}} & = \frac{1 - \ass - \sqrt{\ass(1 - \ass)}}{1 - 2\ass}.
    \end{split}
\end{equation}
A population of discriminators is stable against invasion by defectors provided $b/c$ exceeds
\begin{equation}
    \left( \frac{b}{c} \right)^* = \frac{1 + \sqrt{\dfrac{\ass}{1 - \ass}}}{1 - 2\ass}.
\end{equation}
At exactly this cutoff, however, it becomes vulnerable to invasion by \emph{cooperators}.
There is then a basin of attraction toward a mixed equilibrium of discriminators and cooperators at
\begin{equation}
    f_\text{ALLC} = \frac{b^2(1 - \ass)(1 - 2\ass) - 2 b c (1 - \ass) + c^2}{(b - c) [b(1 - \ass) - c \ass](1 - 2\ass)},
\end{equation}
and so cooperation occurs at rate
\begin{equation}
    f_\text{ALLC} + f_\text{DISC} G = \frac{b^2(1 - \ass)(1 - 2\ass) - 2 b c (1 - \ass) + c^2}{(b - c) [b(1 - \ass) - c \ass](1 - 2\ass)} + \Big( 1 - \frac{b^2(1 - \ass)(1 - 2\ass) - 2 b c (1 - \ass) + c^2}{(b - c) [b(1 - \ass) - c \ass](1 - 2\ass)} \Big) \Big( 1 - \frac{b \ass}{b - c} \Big).
\end{equation}
\subsection{Scoring.} Under Scoring, there is in principle no difference between private and public assessment (at least for $Q = 1$), so we do not consider Scoring further.

\section{Taxation in group-structured populations}

In this section, we expand on the results for group-structured populations; specifically, we consider the fate of tax evaders and bribers when the tax amount $\T$ is reduced due to the fact that, even in the absence of an institution, some cooperation can be maintained.

\subsection{Populations with exchangeable individuals}
We consider first the case where group identity is not explicitly considered in assigning reputations; that is, individuals are not biased toward any particular group---any difference in behavior arises from differences in information transmission.

A taxpaying discriminator has fitness
\begin{equation}
    \begin{split}
        \Pi_\text{taxpayer} & = (1 - \tax) \Pi_\text{institution} + \tax \Pi_\text{group-wise}
        \\
        & = \frac{(b - c) \left(g^\text{in} + \alpha \omega (K-1) [g^\text{in} + \tax (g^\text{out} - g^\text{in})] \right)}{1 + \omega(K-1)},
    \end{split}
    \label{eq:multigroup_tax}
\end{equation}
and a discriminator who shirks the tax, but is detected with probability $\D$ and assigned a bad reputation, has fitness
\begin{equation}
    \Pi_\text{tax evader} = (1 - \D) \Pi_\text{institution}.
\end{equation}
The added term $\tax \Pi_\text{group-wise}$ in Equation \ref{eq:multigroup_tax} means that it is now \emph{less} tempting to dodge one's taxes---because the society did not, in the first place, consent to a high tax rate.
A similar argument applies for defector tax-evaders, whose fitness is
\begin{equation}
    \Pi^\prime_\text{ALLD} = (1 - \delta) \left(b  \frac{G_\text{ALLD} [1 + \alpha \omega (K-1)]}{1 + \omega (K-1)} \right);
\end{equation}
taxpaying discriminators then resist invasion provided $b/c$ exceeds
\begin{equation}
    \left( \frac{b}{c}\right)^* = 1 + \frac{G_\text{ALLD} (1 - \delta)[1 + \alpha \omega (K-1)] }{g_\text{DISC}^\text{in} + \alpha \omega (K-1) [g_\text{DISC}^\text{in} + \tax ( g_\text{DISC}^\text{out} - g_\text{DISC}^\text{in})] - G_\text{ALLD} (1 - \delta)[1 + \alpha \omega (K-1)]}.
    \label{eq:bccond_multigroup_evader}
\end{equation}
In contrast to the case of no group structure, this condition is \emph{always} valid under Stern Judging, i.e., there is \emph{always} some finite $b/c$ that allows taxpaying discriminators to resist invasion by tax-evading defectors.
By rearranging terms, we readily see that the critical value of $b/c$ increases with the tax rate $\tax$ but decreases with the tax evader detection probability $\delta$, as expected.
Furthermore, this is not a very stringent condition.
The worst case scenario corresponds to sending $\alpha \omega (K-1) \to \infty, \tax \to 1, \delta \to 0$, in which case the critical value of $b/c$ is simply $1/(1-2\ass)^2$.
In general, Equation \ref{eq:bccond_multigroup_evader} is \emph{less} stringent than the condition needed for there to be a stable DISC equilibrium in the absence of an institution in the first place (main text equation 25).
In effect, an institution can be maintained ``for free''; if a group-structured population can support cooperation in the absence of an institution, then a tax-collecting institution can raise the rate of overall cooperation without requiring a stricter $b/c$ condition.

Thus, at least in the case of Stern Judging, group structure makes it easier for taxpaying discriminators to persist, by lowering the tax rate.
However, because the institution cannot demand a high tax, the salary of institution members is likewise lower; we have
\begin{equation}
    \salary = N \tax (\Pi_\text{institution} - \Pi_\text{group-wise})/Q.
\end{equation}
The extra $-\Pi_\text{private}$ term means that the minimum amount needed to bribe officials for a good reputation is now lower, so bribing may be a more appealing strategy.
For example, the fitness of ``paranoid'' briber-evaders is
\begin{equation}
    \Pi_\text{ALLD}^\prime = b \left(\frac{1 + \alpha \omega (K-1)}{1 + \omega (K-1)}\right) - N \beta \tax \left( \frac{(b - c) \alpha \omega (K-1)(g_\text{DISC}^\text{in} - g_\text{DISC}^\text{out})}{1 + \omega(K-1)} \right);
\end{equation}
taxpaying discriminators are stable against invasion by this strategy only if $b/c$ exceeds
\begin{equation}
    \begin{split}
    \left( \frac{b}{c} \right)^* & =
    \begin{dcases}
        1 + \frac{1 + \alpha \omega (K-1) }{g_\text{DISC}^\text{in} - 1 + \alpha \omega (K-1) [ g_\text{DISC}^\text{in} - 1 + \tax (g_\text{DISC}^\text{in} - g_\text{DISC}^\text{out})(N \beta - 1)]} & N \beta > (N \beta)^*,
        \\
        \infty & N \beta < (N \beta)^*,
    \end{dcases}
    \\
    \text{with } (N \beta)^* & = 1 + \frac{1}{\tax} \left( \frac{1 - g_\text{DISC}^\text{in}}{g_\text{DISC}^\text{in} - g_\text{DISC}^\text{out}} \right) \left( 1 + \frac{1}{\alpha \omega (K-1)} \right),
    \end{split}
\end{equation}
which immediately allows us to see that the critical value of $N \beta$, above which it is possible to sustain cooperation, is always greater than $1$ and decreases with the tax rate $\tax$, the outgroup interaction probability $\omega$, the number of groups $K$, and the advantage of outgroup interactions $\alpha$.
For Stern Judging, defining $\phi = N \beta - 1$ and $\gamma = \alpha \omega (K-1)$ allows us to simplify this to
\begin{equation}
    \left( \frac{b}{c} \right)^* =
    \begin{dcases}
        1 + \frac{2(1 + \gamma)}{\tax (1 - 2 \ass) \gamma \phi - 2 \ass(1 + \gamma)} & \phi > \frac{2 \ass(1 + \gamma)}{\tax (1 - 2 \ass) \gamma},
        \\
        \infty & \phi < \frac{2 \ass(1 + \gamma)}{\tax (1 - 2 \ass) \gamma}.
    \end{dcases}
\end{equation}

An additional consideration that does not lend itself to an infinitesimal analysis of this kind is that, if an entire group simply unilaterally opts out of paying the tax altogether, institution members' salaries are reduced by a fraction $1/K$; this may create a ``death spiral'' effect that makes it easier for members of other groups to bribe their way into good reputations.
Said group will suffer a bad reputation in the eyes of other groups, but they may compensate for this by becoming more insular--an outcome that is now even more likely to occur, as they have effectively sent $g_\text{out} \to 0$.
Groups with greater wealth or institutional access may likewise coordinate to directly bribe institution members rather than pay the requisite tax.

\subsection{Biased populations}
When reputations are a matter of private judgment, widespread disagreement makes it difficult to reward prosocial behavior.
A straightforward but sinister way to solve this problem is to designate an outgroup whom it is always considered ``good'' to defect against.
Let $\lambda \in (0,1)$ be the size of this group.
In a population consisting entirely of discriminators, the average reputation for the ingroup is
\begin{equation}
    G_\textrm{DISC} = \frac{3 + 4\ass(1 - \lambda) - 2 \lambda - \sqrt{1 + 4(1 - 2\ass)^2(1 - \lambda)\lambda} }{4(1 - 2\ass)(1 - \lambda)}.
\end{equation}
We may reasonably suppose that, in such a situation, the outgroup will reciprocate by refusing to cooperate with the ingroup; thus, multiplying the expression for $G_\textrm{DISC}$ by $1 - \lambda$ yields the total ingroup cooperation rate, and multiplying this by $b - c$ yields the fitness of the ingroup.
The reputation $G_\textrm{DISC}$ is actually maximized for high values of $\lambda$, as it becomes increasingly likely that individuals will be judged based on their interaction with the outgroup--whom everyone agrees is bad--and thus assigned a good reputation.
However, defectors can likewise reap a good reputation by punishing the outgroup, so a higher value of $\lambda$ benefits them, too.
The invasibility of such a population by a rare defector mutant is governed by Equation \ref{eq:bccond_general}, which becomes
\begin{equation}
    \begin{split}
        \left( \frac{b}{c} \right)^* = 1 + \frac{1 - 2 (1 - 2 \ass)^2 \lambda (1 - \lambda) + \sqrt{1 - 4 (1 - 2\ass)^2 \lambda (1 - \lambda)}}{2(1 - 2\ass)^2 \lambda (1 - \lambda)}.
    \end{split}
\end{equation}
It is easily shown that this has a minimum at $\lambda = 1/2$, for which we obtain
\begin{equation}
    \begin{split}
        \left( \frac{b}{c} \right)^* = 1 + \frac{1 + 4 \ass (1 - \ass) + 4 \sqrt{\ass (1 - \ass)}}{(1 - 2\ass)^2}.
    \end{split}
\end{equation}
Surprisingly, this has limit $1$ as $\ass \to 0$.

When an institution is introduced, the reputation of discriminators becomes
\begin{equation}
    \begin{split}
        G_\textrm{DISC} = 1 - \ass,
    \end{split}
\end{equation}
even if the institution exhibits the same bias against the outgroup; that is, there is no dependence on $\lambda$ and thus no benefit to anti-outgroup bias.
As usual, the tax rate a population will pay to sustain such an institution depends on the marginal benefit associated with the institution's existence.
For $\lambda = 1/2$, this is
\begin{equation}
    \begin{split}
        \mathcal{T} & = \Pi_\textrm{institution} - \Pi_\textrm{private}
        \\
        & = (b - c) \frac{1 - \ass (5 - 4 \ass) - \sqrt{\ass (1 - \ass)}}{2(1 - 2 \ass)}.
    \end{split}
\end{equation}
The general solution (see Methods) is an decreasing function of $\lambda$––but the slice of parameter space for which a cooperative equilibrium exists under private assessment likewise decreases (higher $(b/c)^*$) as $\lambda$ exceeds $1/2$.

\bibliographystyle{abbrvnat}
\bibliography{references}